\documentclass[12]{article}
\usepackage{xurl}
\usepackage{authblk}
\usepackage{lmodern}
\usepackage[numbers]{natbib}
\usepackage[a4paper, total={5.5in, 9in}]{geometry}
\usepackage{float}
\usepackage{wrapfig}
\usepackage{amssymb,amsmath}
\usepackage{graphicx}
\usepackage{booktabs}
\usepackage{adjustbox}
\usepackage[table]{xcolor}
\definecolor{Gray}{gray}{0.9}

\providecommand{\keywords}[1]{\textbf{\textit{Keywords---}} #1}
\usepackage[colorlinks=true,linkcolor=blue, citecolor = blue]{hyperref}
\usepackage{longtable}
\usepackage{pdflscape}
\usepackage{setspace}
\usepackage{ulem}

\title{Spatial clusters for demand and supply of childcare services in Italy}
\author[1]{Angela Andreella \thanks{\textbf{Corresponding author}: angela.andreella@unive.it,  San Giobbe, Cannaregio 873, 30121 Venice, Italy}}
\author[2]{Emanuele Aliverti}
\author[1]{Federico Caldura}
\author[1]{Stefano Campostrini}

\affil[1]{\footnotesize Department of Economics, Ca' Foscari University of Venice}
\affil[2]{\footnotesize Department of Statistical Sciences, University of Padova}

\date{ }
\begin{document}
\maketitle
\begin{abstract}
The availability of affordable and high-quality childcare services has become a significant concern in recent years. Such services can facilitate the balance between work and family life, increasing participation in the workforce and promoting gender equality. Furthermore, childcare can also help address the issue of decreasing fertility rates by making it more affordable for parents to have children while maintaining their careers. This is critical, especially for countries that are facing ultralow fertility rates like Italy. The Italian government has included within the recovery and resilience plan financed with Next Generations EU funds an unprecedented investment in order to increase the supply of children's education services and make it more equitably distributed across the country. In this article, we estimate groups of spatial areas with similar structures in terms of coverage (availability of childcare services at the municipality level), public expenditure rates in childcare, as well as other socio-demographic and economic factors, such as female employment, education, and grandparent rates. Our empirical findings confirm how Italy is characterized by a large number of ``sub-regional models'' and how some of these clusters are shared across multiple regions. We provide a preliminary attempt to explain how such patterns are driven by socio-demographic factors and argue that these very different conditions necessitate specific policy decisions. The work highlights the need for regional governance of the children's educational system.
\end{abstract}
\keywords{clustering; childcare services; lasso penalty; multinomial regression; supply and demand; social services; spatial proximity.}

\vspace{.3cm}

\section{Introduction}
In recent years, there has been increasing attention to the availability of childcare services.
Such provisions were established in Italy in $1971$ as ``social services of public interest'', while in the late 1990s, supplementary services for early childhood appeared in the Italian territory. 
These services were instituted as welfare benefits, with the main purpose of supporting parents --- females in particular --- in childcare and participation in the job market. 
Indeed, there is a rich literature on the social impact of childcare services, and economists, educators, and policymakers agree on the relevance of education services for early childhood \citep{plantenga2009provision}; for example, there is common agreement on how such facilities promote gender employment equality \citep{landivar2021states, addabbo2012allocation, del2002effect, chiuri2000quality, meyers1999public} and the related fight against an aging population \citep{luci2013impact, haan2011can, del2002effect}. In addition, childcare services have a notable educational and social impact on a child's cognitive development, particularly when the service offers high-quality education \citep{felfe2018does,van2018children,felfe2015can,brilli2016does}.
Moreover, several studies indicate that decreasing service prices and increasing available places impact the families' decision to send the child to daycare, improving the mothers' work participation. 
We emphasize here that our work aims to explore another point of view. Instead of focusing on the impact of welfare policies on the population's social and economic aspects, this paper wants to inspect the complex and multidimensional feature of the supply/demand combination of education services in Italy, jointly with the variations in the socio-demographic and economic structure.

Relying on a national perspective, it is well known that Italy is characterized by a low child coverage ratio, measured as the number of available places in childcare structures divided by the number of toddlers (children under $3$ years) and labor force participation is particularly low \citep{ISTAT}.
This situation has stimulated an intense discussion about the role of these services, and there is a consensus that the Italian children's education system must be improved. From a practical point of view, there are several difficulties for the supply system to respond quickly to changes in socio-cultural frameworks and the concrete need for educational services with high-quality standards throughout the country. 
Several analyses developed thanks to the collaboration between the Department of Family Policies, ISTAT (Italian national institute of Statistics), and the Ca' Foscari University of Venice - Department of Economics showed the persistence of important issues \citep{ISTAT}. Firstly, a structural lack of services has emerged, despite the potentially large demand. Second, the spatial distribution of childcare services is deeply irregular in the national territory.

Focusing on the level of service coverage, \cite{ISTAT} reported a first substantial supply variability. 
Notably, coverage across the Italian territory is below the European average and the European Union target, which is fixed at $33\%$. There is also significant heterogeneity across the Italian territory, with South Italy and the Islands (Sicilia and Sardegna) being substantially disadvantaged. 
For example, focusing on the year $2019$, North-East and Central Italy confirm the level of coverage above the European target (with $34.5\%$ and $35.3\%$ respectively), and North-West Italy is getting closer to the target ($31.4\%$), while the South and the islands are still far from the target (with $14.5\%$ and $15.75\%$ respectively). 
Such variability in the supply is also evident in terms of the type of services (i.e., public and private ones) \citep{ISTAT}. 
In fact, South Italy has a percentage of public childcare services equal to $41.1\%$, while in the North-East, it equals $55.1\%$ (above the national mean equals $49.1\%$). 

Most importantly, \cite{ISTAT} noted an important variability at a much finer spatial level than those canonical macro areas, i.e., between the Italian regions and also sub-regionally. 
The greatest coverage is found in the larger municipalities, like Bologna, Firenze, Roma, Milano, and Venezia (with coverage equals $\approx 40\%$). 
For example, the coverage in Torino equals $\approx 35\%$ if the metropolitan area is considered while it equals  $\approx 25\%$ outside the suburban area. 
Unfortunately, the gap between the North-Central and the South does not diminish focusing on the urban areas. 
Regarding the supply side, \cite{ISTAT} observed a notable difference between North and South concerning the per-capita public expenditure. Optimal situations are in regions like Trentino (considered an autonomous region), Valle d'Aosta, Emilia Romagna, Lazio, and Toscana, while suboptimal conditions are observed in Calabria, Campania, Puglia, and Basilicata. 
Such variability within the same macro-area is also consistent if the percentage of public services is analyzed. 
For example, the large percentage of public services in North-East Italy is mainly due to the optimal situation in Emilia Romagna ($71.2\%$), Bolzano ($54.1\%$), and Trento ($76\%$). The remaining northeast regions (i.e., Veneto, Friuli-Venezia Giulia) have percentages significantly below the national average \cite{ISTAT}.

\cite{ISTAT} also analyzed some aspects of the demand side of childcare services in Italy through the sample survey ``Aspects of daily life'' \citep{ISTATmultiscopo}. In this survey, $25000$ households distributed in about $800$ Italian municipalities are interviewed about school, work, family, and relationship life, housing and area of living, leisure time, political and social participation, health, and lifestyles. 
\cite{ISTAT} found that most of the children, who make use of these services, have working parents ($\approx 60\%$), the father with a high level of education (i.e., greater than high school graduation) ($\approx 41,6\%$), and the family without economic problems ($\approx 30\%$). Finally, \cite{ISTAT} observed that one of the main reasons for the non-participation in childcare services is the geographical distance (i.e., the child resides in a suburban municipality in the metropolitan area), the fact that a family member can take care of the child, and the childcare service cost. The existing literature also confirms these aspects (at least in several American and European countries) \citep{ anderson1999child, baker2008universal, cornelissen2018benefits}.

Although these differences have been extensively documented, the motivations of this variability are still unexplained, particularly in terms of the socio-demographic structure of the population.
Motivated by these questions, this work aims to analyze the childcare services' supply and demand structure to understand how they differ within the Italian system. This study wants then highlight how educational services do not follow a ``regional pattern" but are instead affected by strong sub-regional variability driven by socio-demographic factors. 

The analysis is developed at the level of administrative unit (ATS) within the region and the province.
From a theoretical and political perspective, the choice to focus on this micro-scale level to explore the intra-regional variability of childcare services can be motivated by various factors. 
In Italy, the educational offer for children from zero to six years old consists of two segments that have been characterized by very different developments, even if they are considered today within a single integrated offer of educational services. 
Indeed, childcare services for toddlers have historically been included in the sphere of social services instead of educational services (such as primary schools).
For this reason, they have long been the subject of decentralized decisions with a substantial absence of central governance. 
In addition to a poorly controlled and uneven proliferation of detailed regulatory acts and interventions at local government levels, this development has brought evident territorial differences in providing childcare services \citep{ISTAT}.

The ATS are identified by regions to support or replace the municipalities in managing social interventions and services (Italian law d.lgs328/2000) with the objective of reducing these micro-territorial differences in social interventions. They are formally defined as inter-municipal aggregations that handle social programming and often intersect with the scheduling of childhood education.
Furthermore, ATS can be considered a plausible territorial dimension to obtain a description more in line with the reality of accessibility to educational services, taking into account their distribution in areas larger than municipal ones and, simultaneously, bringing out any specific differences between certain areas, i.e., within regions and provinces. In fact, widespread and homogeneous distribution of childcare services at the municipal level is rare to be found. In most of them, the supply of services is concentrated in some municipalities more than in others, and it is well known that several families bring their children to a different municipality due to the lack of available places in the municipality of residence \cite{aliverti2021}.

The main objective of this work is to propose a suitable statistical approach to find clusters of ATS that are similar in terms of supply and demand structure, not necessarily within the same region. 
We focus on $2019$ data on childcare services from the ISTAT survey ``Survey of child care and early childhood supplementary services'' \citep{ISTATsurvey} that describes the supply side and its structure. Finally, relevant cultural and social features that could characterize the structure of the demand \citep{ISTAT}, such as the fertility rate, the presence of family support (e.g., grandparents), and the parents' work type, are included in the analysis using data from the $2019$ permanent census \citep{ISTATcensimento}.

The analysis is divided into two steps. First, a clustering method with geographical constraints is applied to provide insights into the variability of the supply and demand combination within the Italian territory (i.e., highlight sub-regional childcare welfare models). 
We use a regionalization method called SKATER (Spatial ’K’luster Analysis by Tree Edge Removal, \citep{assunccao2006efficient}), which considers the spatial proximity of ATS in the formation of clusters.
A distance-based clustering algorithm allows highlighting clusters of ATS that also belong to different regions, while the additional spatial constraints facilitate the estimation of smooth clusters across neighbors ATS, improving the interpretability of the estimated structure. As a second step, we fit a penalized multinomial model \citep{friedman2010regularization} to link the clustering indicator with supply and demand covariates, estimating the influence of such factors on the probability of belonging to a specific cluster.

The paper is organized as follows. Section~\ref{data} introduces the variables analyzed coming from the two datasets described before, i.e., the ISTAT survey ``Survey of child care and early childhood supplementary services'' \citep{ISTATsurvey} and the permanent census $2019$, \citep{ISTATcensimento}. In particular, Subsection~\ref{descriptive} shows some descriptive statistics, while Subsection~\ref{method} introduces the SKATER regionalization method \citep{assunccao2006efficient}. Finally, Section~\ref{results} outlines the results of the spatial clustering analysis and penalized multinomial model. 

\section{Data description and statistical analysis}\label{data}
In this work, two data sources are used to capture the complex and multidimensional structure of the demand and supply of Italian childcare services. We focused on a single year ($2019$) to have available and up-to-date data and to provide a snapshot of the childcare services' spatial variability rather than making a temporal comparison. 

As outlined in the Introduction, we decided to consider the ATS as spatial statistical units. The Italian territory is divided into $20$ regions divided into provinces, which are, in turn, divided into municipalities. For example, the Veneto region is divided into $7$ provinces and $563$ municipalities. Table \ref{tab:ATS} shows the number of ATS for each Italian region; current data are composed of $n=606$ ATS.

\begin{table}[h]
    \centering
\begin{tabular}{lr}
\toprule
\textbf{Region} & \textbf{Number of ATS}\\
\midrule
Piemonte & 49\\
Valle d'Aosta/Vallée d'Aoste & 5\\
Lombardia & 88\\
Trentino-Alto Adige/Südtirol & 17\\
Veneto & 21\\
Friuli-Venezia Giulia & 18\\
Liguria & 18\\
Emilia-Romagna & 38\\
Toscana & 26\\
Umbria & 12\\
Marche & 23\\
Lazio & 37\\
Abruzzo & 24\\
Molise & 7\\
Campania & 57\\
Puglia & 45\\
Basilicata & 9\\
Calabria & 32\\
Sicilia & 55\\
Sardegna & 25\\
\bottomrule
\end{tabular}
    \caption{Number of ATS for each Italian region.}
    \label{tab:ATS}
\end{table}

The supply side of childcare services is described using data from the ISTAT survey ``Survey of child care and early childhood supplementary services'' \citep{ISTATsurvey}, 
while the demand side by data from the $2019$, permanent census of $2019$ \citep{ISTATcensimento}. 
Table \ref{tab:var} briefly describes the variables used in the analysis. The first two (i.e., coverage and per capita public expenditure rate) represent the supply side, and the remaining variables describe the demand side. The latter variables were chosen to define possible solutions that can be roughly considered an alternative to childcare services (e.g., babysitters, grandparents, extended families) and to outline the socio-economic characteristics of the population of each ATS.

\begin{table}
\begin{adjustbox}{max width=\textwidth,center}
\begin{tabular}{ll}
\toprule
\textbf{Variable} & \textbf{Description}\\
\midrule
Coverage & Number of day-care places/Number of resident children \\
& between $0$-$2$ years old\\
Per capita public expenditure rate & (calculated on the resident population between $0$-$2$ years old)\\
\midrule
Female employment rate & Number of resident females working/number of resident\\
& females between $20$-$64$ years old\\
Female house rate & Number of resident females not working at home/number \\ & of resident children between $0$-$2$ years old\\
Commuter rate & Number of commuters for work outside the municipality\\
&/number of workers\\
Male educational qualification rate & 
Number of males with a degree higher than high\\
& school degree/number of males over $20$ years of age \\
Female educational qualification rate & 
Number of females with a degree higher than high\\
& school degree/number of females over $20$ years of age \\
Foreign rate & Number of foreign residents/number of residents\\ 
Grandparent rate &  Number of resident retired persons/number of resident\\ &children between $0$-$2$ years old\\
Babysitter rate & Number of resident females not working (studying) between \\ &$15$-$25$ years old/number of resident children between $0$-$2$ years old\\
Number of members in the household & \\
Fertility rate & Number of resident children with age $0$/ number of resident \\ &females between $15$-$49$ years old.\\
Social and Material Vulnerability Index & It comprises seven different indicators capturing household\\  (IVSM) & size, education level, employment status, welfare distress, \\ &and severe crowding condition in the housing situation. \\ &For a more detailed description, please refer to \cite{IVSM}\\

\bottomrule
\end{tabular}
\end{adjustbox}
    \caption{List of the variables analyzed. The first two variables describe the supply side of childcare services, while the remaining outline the demand side.}
    \label{tab:var}
\end{table}

\subsection{Descriptive statistics}\label{descriptive}
We propose here some exploratory plots to show the data variability, while the remaining are placed in Appendix \ref{app}. 
Figure \ref{fig:descriptive} show the relationship between the coverage and the female employment rate (left figure), grandparents rate (center figure), and the Social and Material Vulnerability Index (IVSM) (right figure) divided by Italian macro areas, i.e., North-East (Trentino-Alto Adige, Veneto, Friuli-Venezia Giulia, Emilia-Romagna), North-West (Valle d'Aosta, Liguria, Lombardia, Piemonte), Center (Toscana, Umbria, Marche, Lazio), South (Abruzzo, Molise, Campania, Puglia, Basilicata, Calabria) and Islands (Sicilia, Sardegna). The black lines represent the corresponding fitted local polynomial regression. 

\begin{figure}[!htb]
     \centering
     \includegraphics[width=\linewidth]{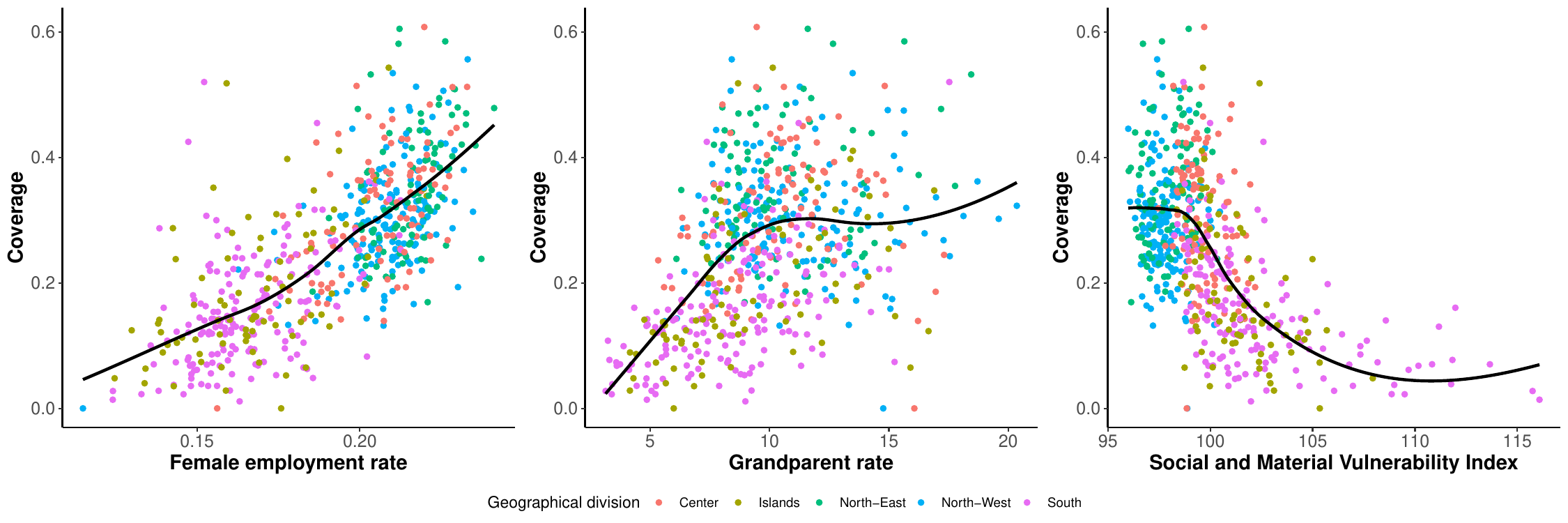}
   \caption{Scatter-plots between coverage and the female employment rate (left figure), grandparents rate (center figure), and the Social and Material Vulnerability Index (IVSM) (right figure) divided by macro areas (Center, Islands, North-East, North-West, and South). The black lines represent the fitted local polynomial regression.}\label{fig:descriptive}
\end{figure}

The left panel of Figure~\ref{fig:descriptive} highlights a positive relationship between coverage and female employment rate, as suggested by the existing literature \citep{landivar2021states, chiuri2000quality, meyers1999public}. In particular, northern areas have higher coverage and participation of females at work, in stark contrast to the situation in southern Italy \citep{ISTAT}.
In the middle panel of Figure~\ref{fig:descriptive}, we can observe a positive relationship between coverage and grandparent rate, although the overall strength is smaller. However, the grandparent rate should be considered with caution. It was built to represent a viable alternative to daycare, but due to a lack of data, the index was built by considering the total population of retirees. 
Therefore, it does not exclude the older population, who can hardly support children.
Moreover, this index is influenced by the demographic characteristics of the region itself: It will have very large values in regions characterized by an older population, such as Liguria and Sardegna. 
Finally, it is interesting to note the relationship between coverage and the IVSM index, reported in the right panel of Figure~\ref{fig:descriptive}, where it is evident that Italian areas with low social and economic levels are deprived of children's educational welfare. 
As found by \cite{ISTAT} and numerous literature \citep{cornelissen2018benefits, baker2008universal, anderson1999child}, childcare services are often affordable only for families with specific economic and social characteristics, thus causing a critical social gap within the Italian territory.

\begin{figure}[!htb]
   \begin{minipage}{0.45\textwidth}
     \centering
     \includegraphics[width=\linewidth]{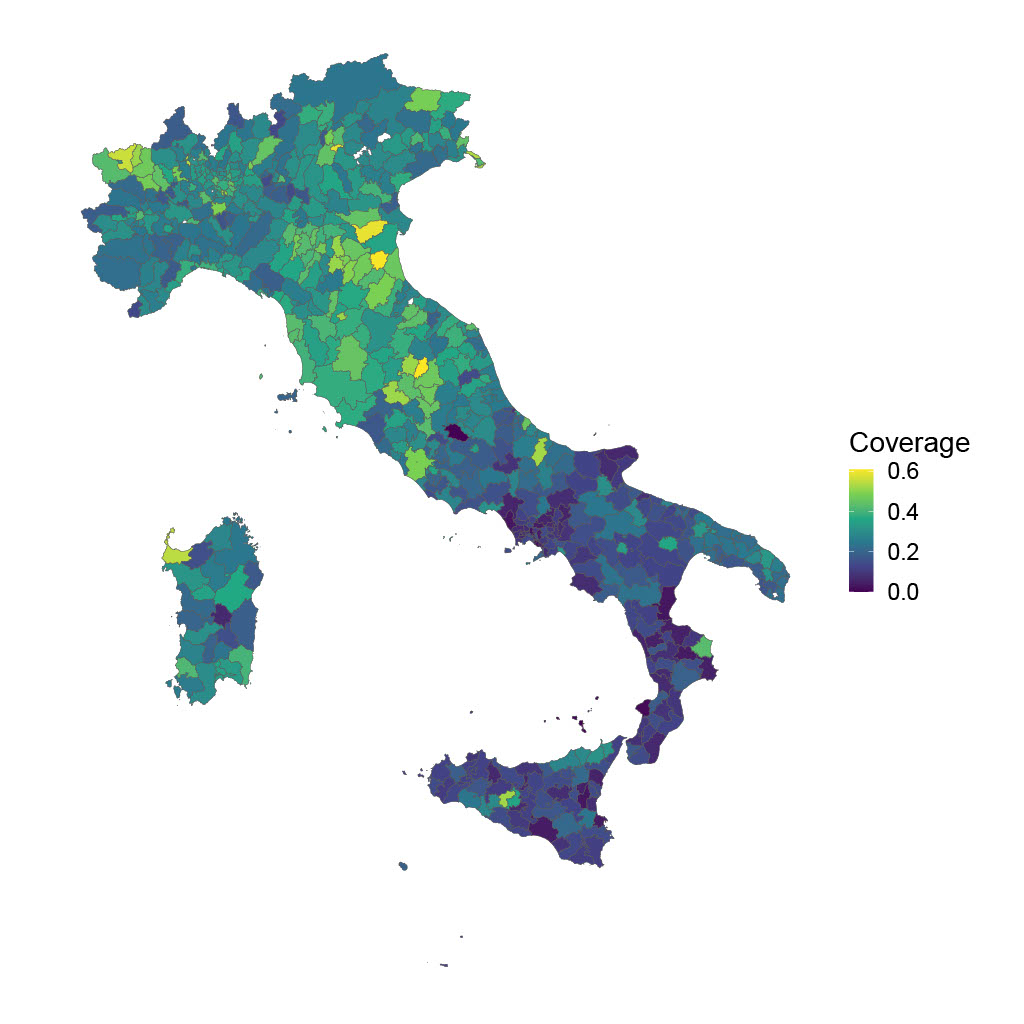}
   \end{minipage}\hfill
    \begin{minipage}{0.45\textwidth}
     \centering
     \includegraphics[width=\linewidth]{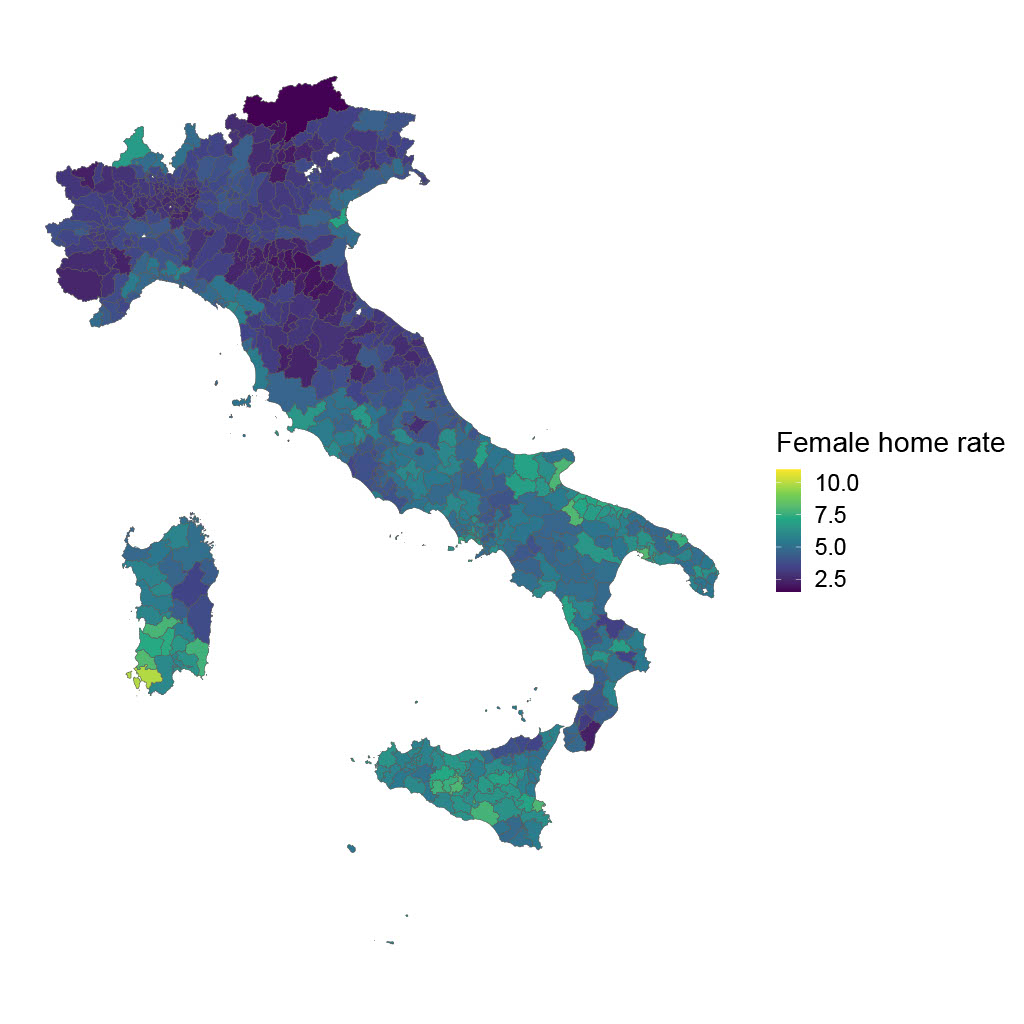}
   \end{minipage}\vfill
    \begin{minipage}{0.45\textwidth}
     \centering
     \includegraphics[width=\linewidth]{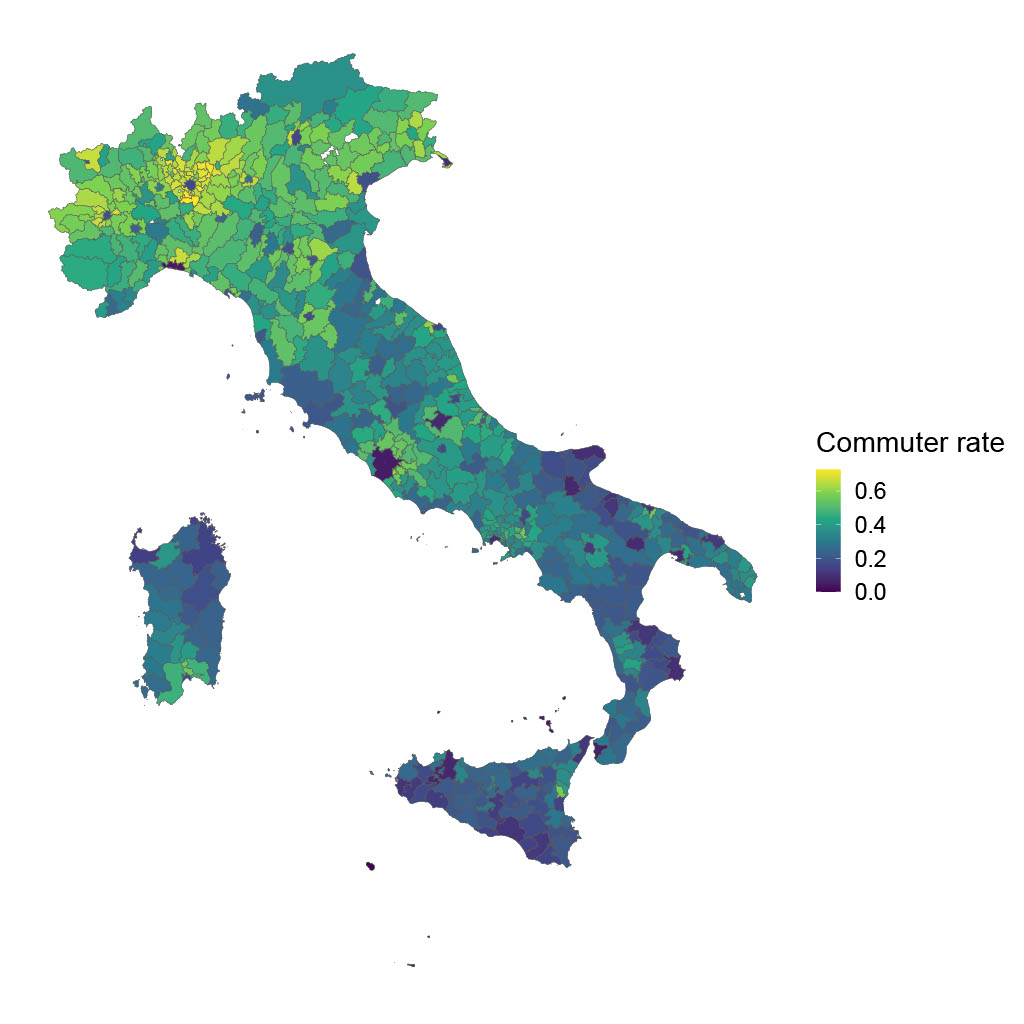}
   \end{minipage}\hfill
    \begin{minipage}{0.45\textwidth}
     \centering
     \includegraphics[width=\linewidth]{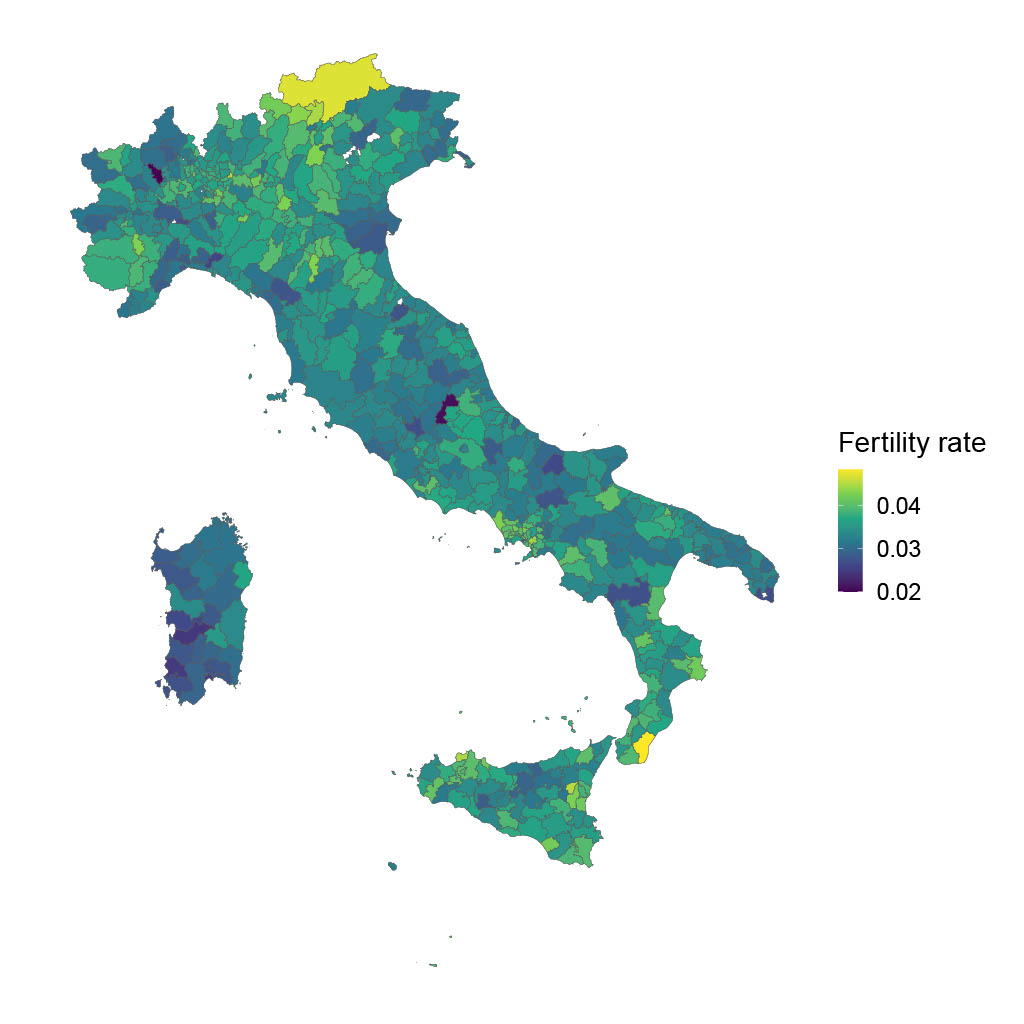}
   \end{minipage}
   \caption{Geographical map representation of the coverage rate (left top figure), female home rate (right top figure), commuter rate (left bottom figure) and fertility rate (right bottom figure) at ATS spatial level.}\label{fig:descriptive_map}
\end{figure}

The four panels of Figure~\ref{fig:descriptive_map} show the coverage (top-left), female home (top-right), commuter (bottom-left), and fertility  (bottom-right) rates at ATS spatial level. 
Looking at the top part of Figure~\ref{fig:descriptive_map}, the gap between north and south in coverage and females at home rate is clear.
Northern Italy is characterized by a proportion of females at home less than southern Italy and a coverage rate greater than southern Italy, as previously seen in Figure~\ref{fig:descriptive}; clearly, the rate of females at home is directly related to the rate of female work analyzed in Figure \ref{fig:descriptive}. 
However, Figure~\ref{fig:descriptive_map} provides further insights into the strong variability observed in these two rates at the regional and sub-regional levels, discussed in the Introduction.
For example, some ATS of the north (e.g., in Emilia Romagna) report larger coverage than the nearby ATS, and similar patterns are observed in Sardegna and Abruzzo. In contrast, the rate of females at home seems more spatially uniform.

Similarly, the bottom-left part of Figure~\ref{fig:descriptive_map} indicated that the commuting rate is greater around large metropolitan areas. 
This index is important to consider because it suggests families' possible displacement and, therefore, a different need for educational services regarding the lease. Commuter parents will probably choose a daycare near work, for example, in another ATS rather than in the place of residence or near the place of residence \citep{aliverti2021}. It is well-known in the literature that the probability of a female taking a job decreases as the work travel time increases. Females are more likely to take flexible jobs close to home to balance work and private life more easily \cite{farre2020commuting, borghorst2021commuting}. This again leads to gender inequality, which policymakers need to consider by, for example, creating daycare places in those areas marked by commuting workers.
Finally, the right part of Figure \ref{fig:descriptive_map} represents another critical aspect of the demand for child education services: the fertility rate. 
Focusing on the national scale, there does not seem to be a significant gap between the North and South. However, we can point out that the fertility rate of Südtirol, known for its innovative family welfare (i.e., social farming), is among the highest in Italy \citep{gramm2020farms}. 
While this fact suggests that an efficient social policy can help to increase the birth rate of a population \citep{bauernschuster2013does}, at the same time, it is worth noting that the relationship between childcare and fertility is likely to be complex and influenced by many more factors such as personal preferences, economic considerations, and cultural norms \citep{ISTAT}.

Overall, there is much evidence of regional and sub-regional variability in the supply and demand of childcare services. 
Clearly, these descriptive statistics do not provide sufficient information on the behind complex relationships, and more sophisticated models are necessary to achieve this aim.

\subsection{Spatial clustering}\label{method}

In order to highlight groups of similar ATS, with respect to the observed characteristics outlined in Section~\ref{descriptive}, we propose a cluster analysis.
Traditional clustering approaches (e.g., $k$-means \citep{macqueen1967classification} and hierarchical clustering \citep{johnson1967hierarchical}) minimize the dissimilarity within-group and maximize the dissimilarity between groups. 
Often, such similarities are expressed in terms of distances. However, these methods do not consider spatial contiguity, which is an important aspect to be accounted for in the modeling of territorial services; indeed, we expect that spatially contiguous ATS might share similar levels of services, and this information should be included in the modeling approach. Due to this, we will consider clustering algorithms for spatial data that are able to exploit the available spatial information in constructing the clusters; refer to \cite{xu2015comprehensive} for a complete review of clustering methods.
A possible method for defining homogeneous spatial clusters following certain variables is regionalization \citep{wise2001providing}.
This technique of spatial statistics is used in various disciplines, and it has been used to capture environmental \citep{bernetti2010minimizing}, socio-economic \citep{openshaw1995algorithms}, epidemiological \citep{haining1994constructing} and climatic \citep{fovell1993climate} spatial clusters.
Regionalization aims to maximize the within-cluster similarity and minimize the between-cluster similarity taking into account the spatial information of the data. These methods are formulated as constrained optimization problems, where the constraint is based on the spatial contiguity of the spatial units. 

Let us consider the $p = 13$ variables of Table \ref{data} measured for each spatial unit (ATS), i.e., $x_i = (x_{i\,1}, \dots, x_{i\,13})$ with $i = 1, \dots, 606$. 
The regionalization approaches aim to find $k$ groups of spatial units, where units in the same group are characterized by similar values of $x$ and units in different groups by different values. Each spatial area is described as a polygon with vertices and boundaries. The spatial constraint is based on the spatial neighborhood structure defined by geographical adjacency: the spatial areas share at least one boundary or vertex (i.e., queen approach \citep{bondy1976graph}). Summarizing the spatial units by their centroids, the spatial constraint can be relaxed. We then define that two spatial areas $i$ and $j$ are spatially contiguous if and only if the distance $d_{ij}$ (e.g., euclidean distance) is less than a fixed maximum distance $r \in \mathbb{R}^+$. The elements of the adjacency matrix $\textbf{W}= [w_{ij}]$ with dimensions $606 \times 606$ are then defined as
\begin{equation}\label{eq:W}
  w_{ij} =  \begin{cases}
 1 & \text{if } d_{ij} \le r \\
 0 & \text{if } d_{ij} > r \\
    \end{cases}
\end{equation}
and is used to construct the graph. Here, the distance is computed considering the latitude and longitude of the centroids for each spatial unit $i$. Because we are considering relatively small areas with fairly regular shapes, the centroid represents a reasonable summary measure of the spatial area described by the polygon.

Leveraging on the SKATER algorithm \cite{assunccao2006efficient}, the spatial units summarized by their centroids can be modeled as a node in an undirected graph \citep{tutte2001graph} represented via the adjacency matrix of Equation \eqref{eq:W}. Therefore, the spatial constraint of the SKATER algorithm simply enters when the initial graph is constructed, i.e., when the adjacency matrix $W$ defined in Equation \eqref{eq:W} is determined. The spatial clusters are then defined as connected subgraphs that minimize the within-cluster heterogeneity, computed by associating a cost or weight $c_{ij}$ to each edge that connects the node (i.e., spatial unit) $i$ and $j$. 
These weights are based on the squared Euclidean distance between locations $i$ and $j$ with respect to their attribute vectors $x_i$ and $x_j$. The within-clusters similarity is then described as the Euclidean squared distance between the location attributes in cluster $k$ and the cluster means of these attributes.
The SKATER algorithm is based on the minimum spanning tree approach to reduce the graph complexity. After constructing the corresponding minimum spanning tree, the SKATER algorithm prunes the tree for the desired number of clusters to minimize the within-clusters variability; refer to \citep{assunccao2006efficient} for further details.

After clustering the spatial units, a multinomial regression model with $\ell_1$ penalization on the coefficients is fitted on the clustering indicators against the socio-demographic factor. Such an approach allows explicitly linking the probabilities of being in a specified cluster with covariates described in Table \ref{tab:var}. 
The regularization  notably improves interpretation, as few variables are selected for the membership to each cluster.

\section{Results}\label{results}

\begin{figure}
    \centering
    \includegraphics[width=.7\linewidth]{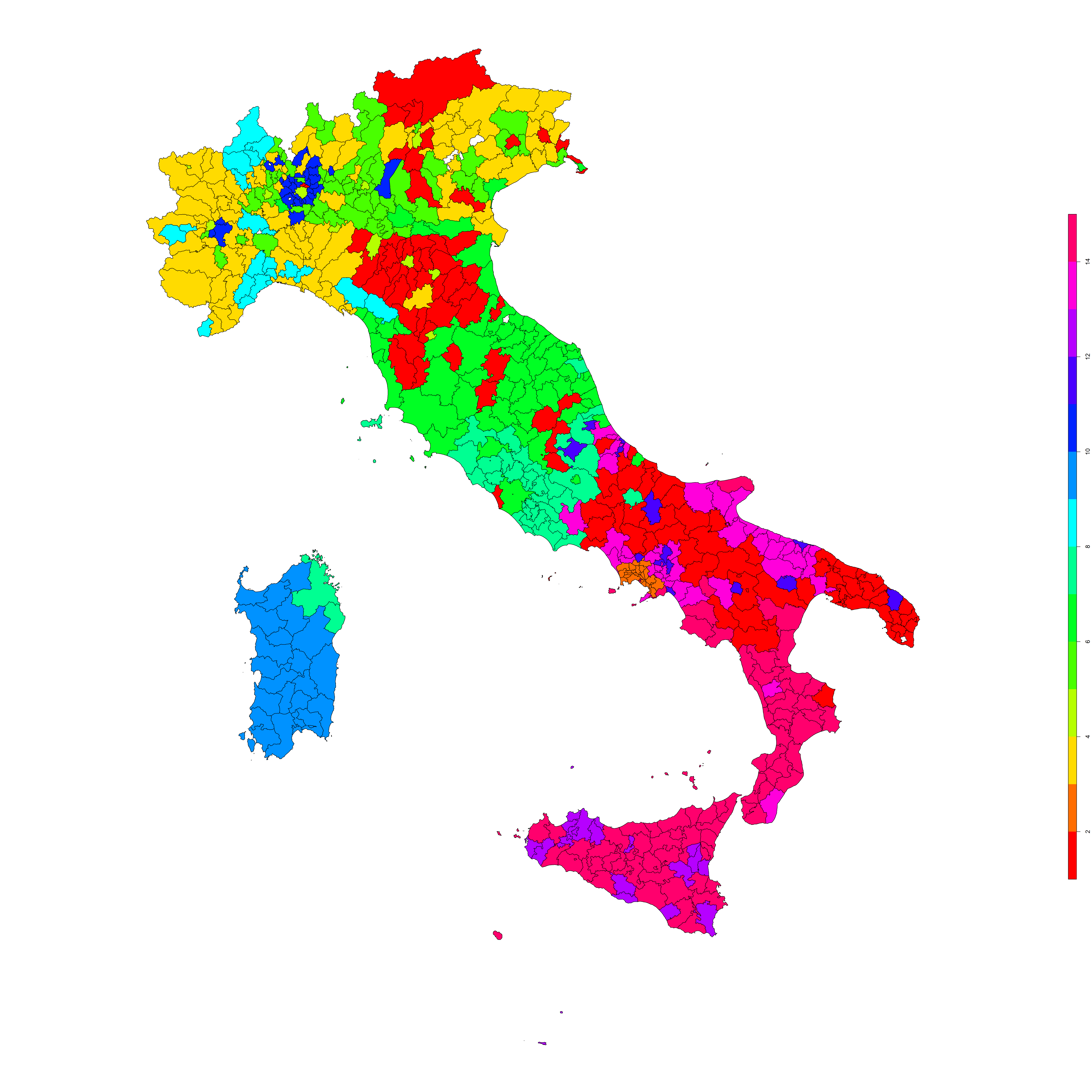}
    \caption{Geographical representation of the $15$ clusters created by the SKATER algorithm with flexible spatial constrain.}
    \label{fig:global_map}
\end{figure}

\begin{figure}
    \centering
    \includegraphics[width=.7\linewidth]{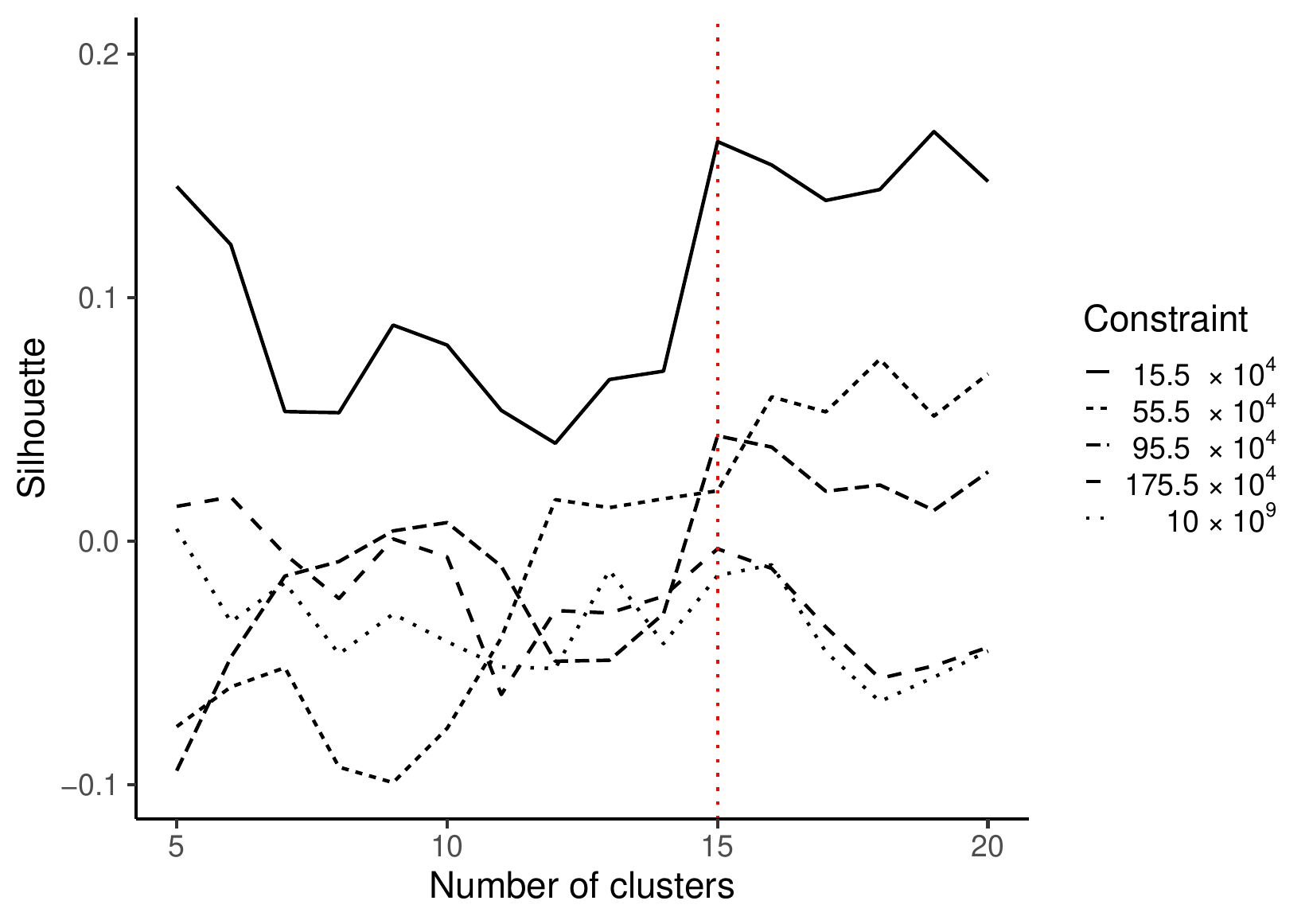}
    \caption{Silhouette values considering several number of clusters $k$ and several values of spatial constraint $r$. The black solid line corresponds to the spatial constraint presented in this paper.}
    \label{fig:sil}
\end{figure}

Figure \ref{fig:global_map} describe the clusters created by the SKATER algorithm imposing $k = 15$, i.e., the number of clusters. We decided to consider a large number of clusters to explore better the variability of the combination between the demand and the supply characteristics of childcare services across the Italian territory. In addition, analyzing the silhouette measure \citep{hastie2009elements} $k=15$ is a reasonable choice. 
Figure \ref{fig:sil} shows the silhouette value across several $k$ values (i.e., $k \in \{5 \dots, 20\}$ and $r$ spatial constraints (i.e., $r \in \{ 15.5 \times 10^4, 55.5 \times 10^4, 95.5 \times 10^4, 175.5 \times 10^4, 10 \times 10^9\}$). The black solid line is the one considered in our analysis related to the minimum value $r$ to have a connected graph. As $r$ increases, the spatial constraint is then relaxed. We can note how our case (i.e., $r = 15.5 \times 10^4$ ) outperforms in terms of silhouette the other situations in all cases.

\begin{figure}[!htb]
\centering
   \begin{minipage}{0.33\textwidth}
     \centering
     \textbf{Cluster 1 [Group 1]}\par\medskip
     \includegraphics[width=.7\linewidth]{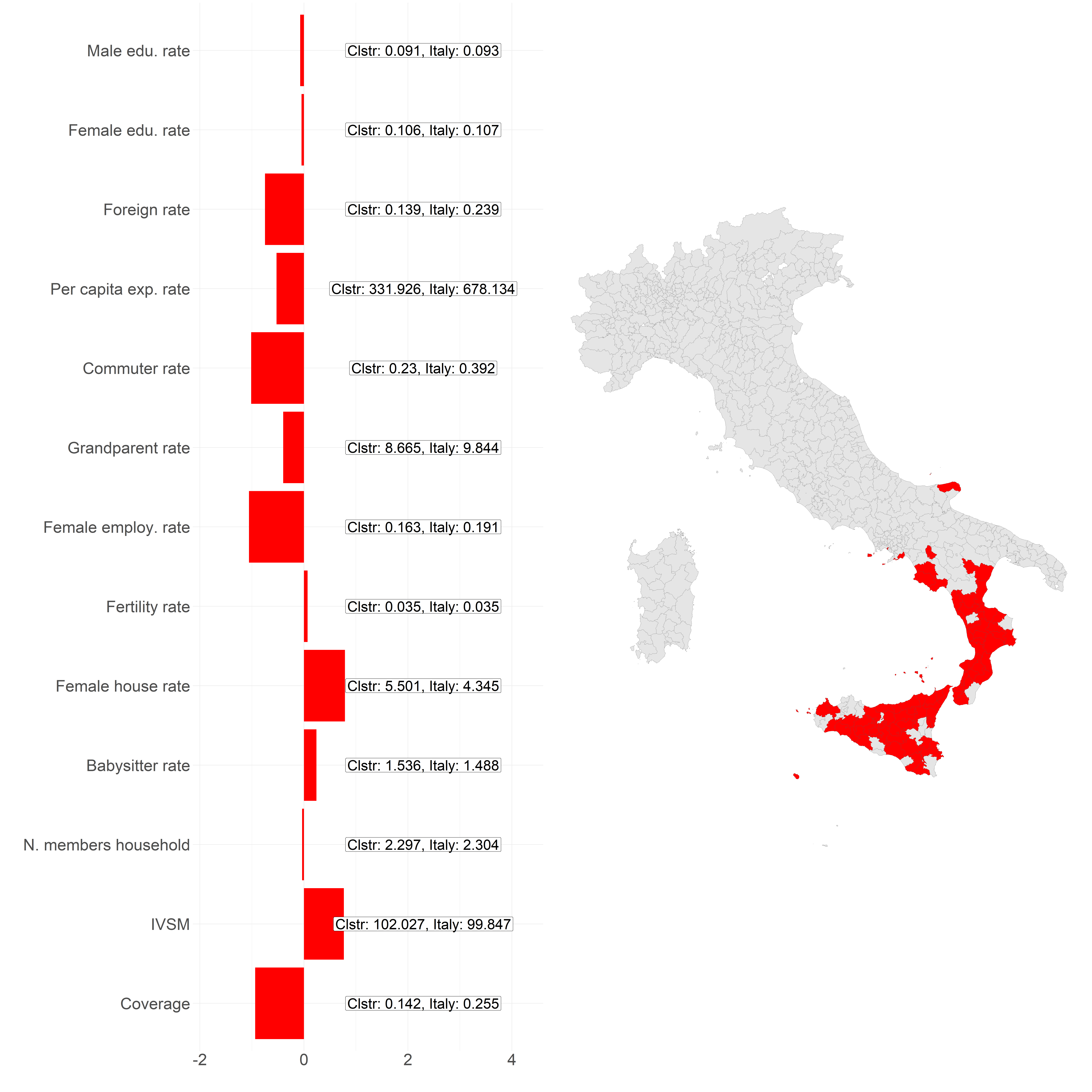}
   \end{minipage}\hfill
    \begin{minipage}{0.33\textwidth}
     \centering
     \textbf{Cluster 2 [Group 1]}\par\medskip
     \includegraphics[width=.7\linewidth]{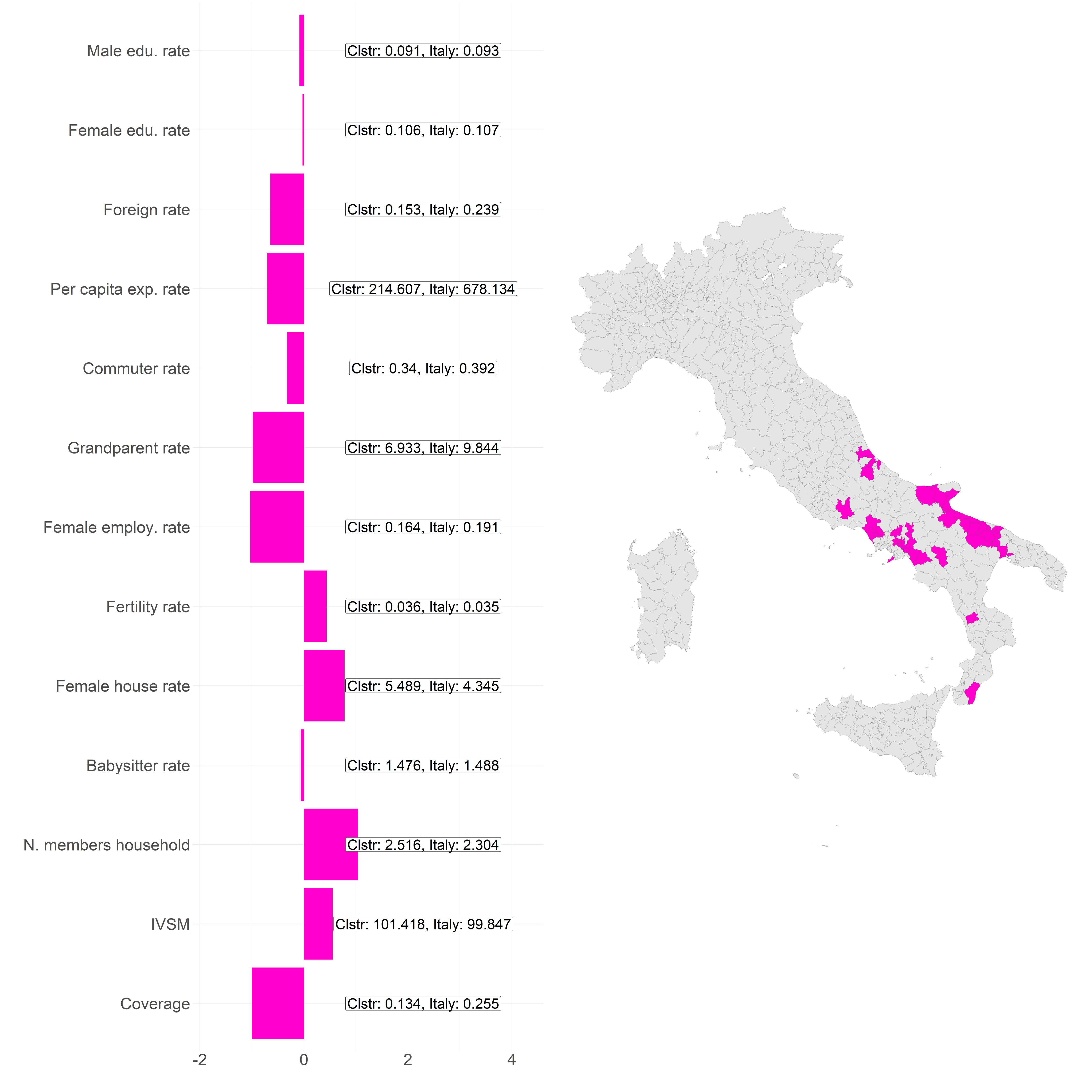}
   \end{minipage}\hfill
   \begin{minipage}{0.33\textwidth}
     \centering
     \textbf{Cluster 3 [Group 1]}\par\medskip
     \includegraphics[width=.7\linewidth]{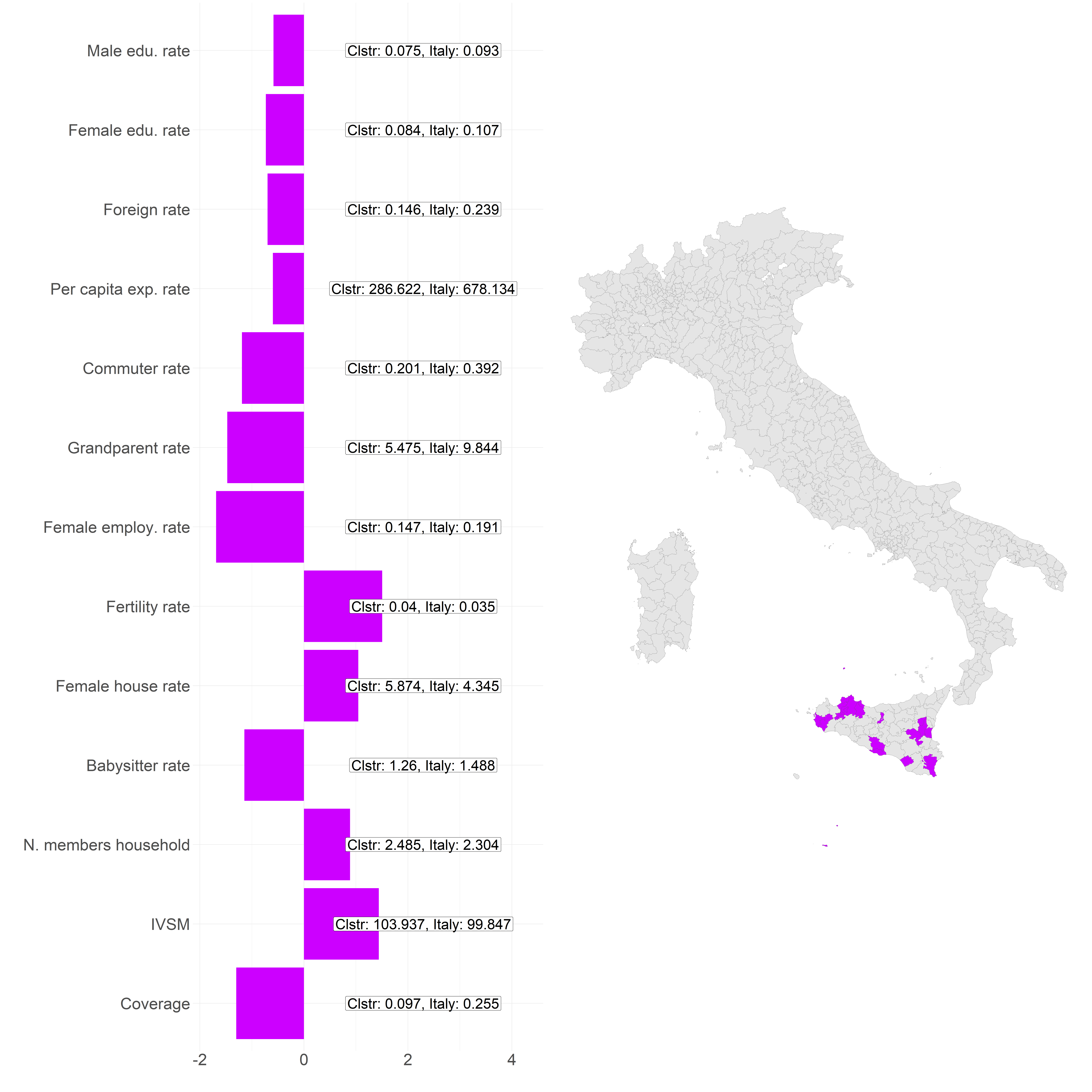}

   \end{minipage}\vfill
      \begin{minipage}{0.33\textwidth}
     \centering
     \textbf{Cluster 4 [Group 1]}\par\medskip
     \includegraphics[width=.7\linewidth]{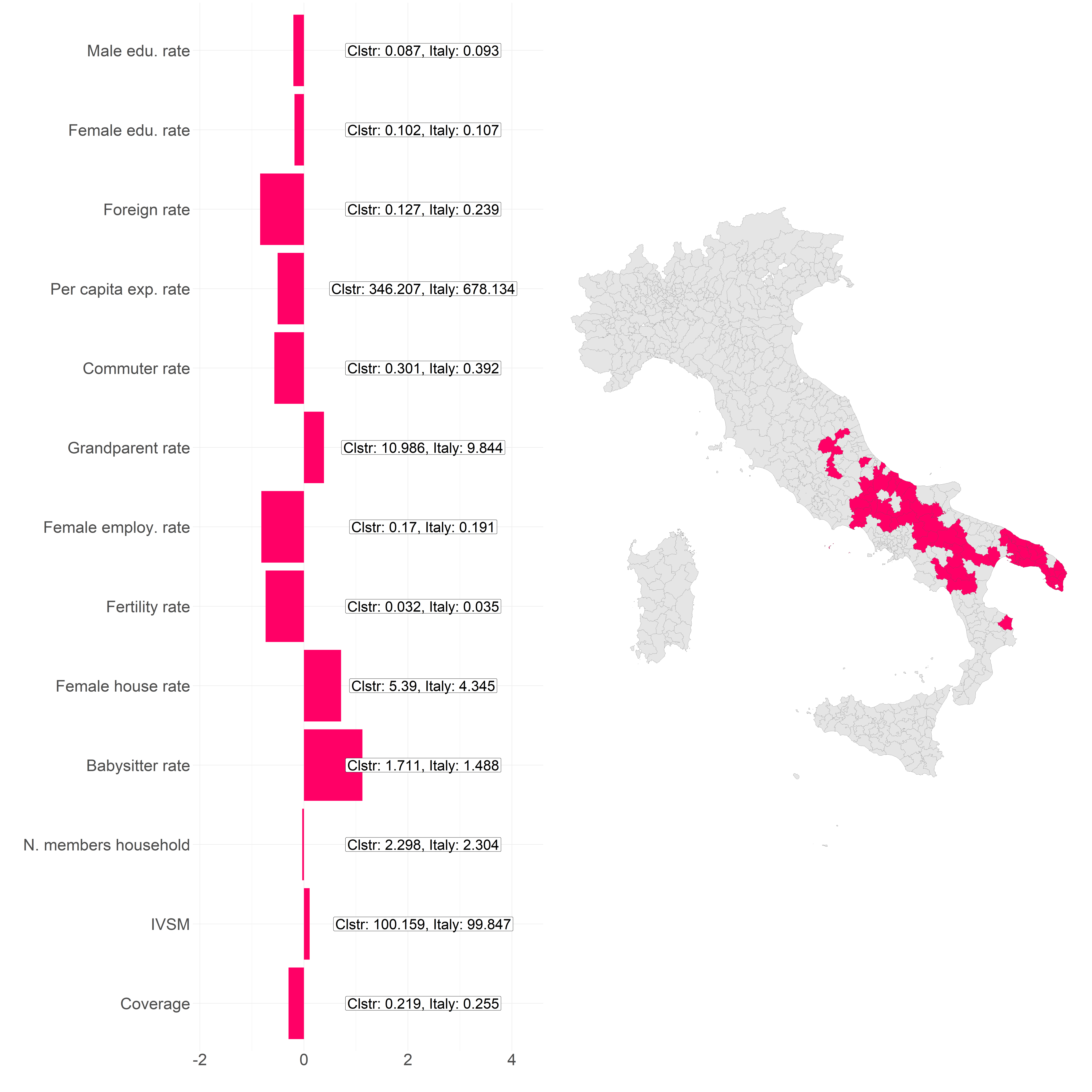}
   \end{minipage}\hfill
    \begin{minipage}{0.33\textwidth}
     \centering
     \textbf{Cluster 5 [Group 2]}\par\medskip
     \includegraphics[width=.7\linewidth]{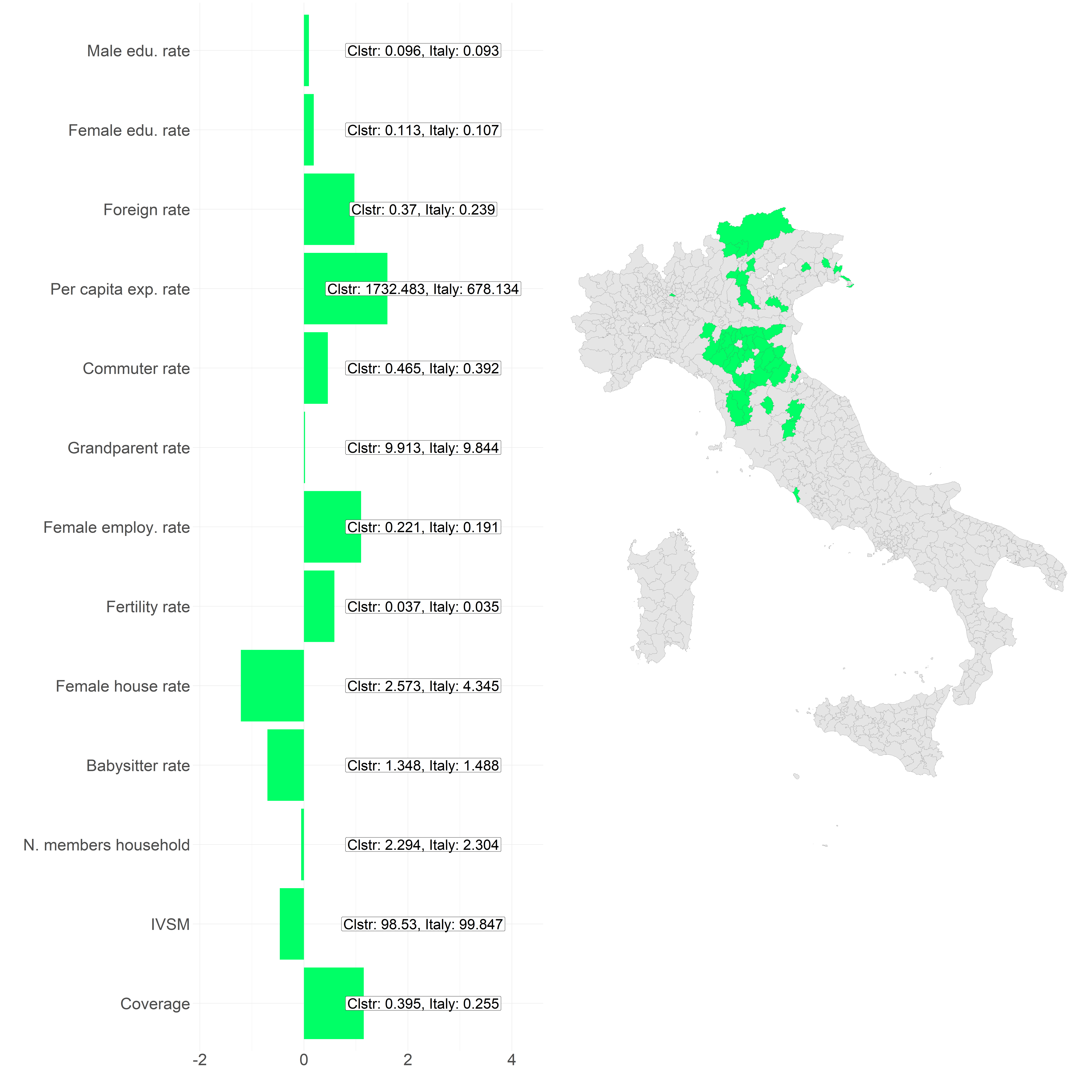}
   \end{minipage}\hfill
   \begin{minipage}{0.33\textwidth}
     \centering
     \textbf{Cluster 6 [Group 2]}\par\medskip
     \includegraphics[width=.7\linewidth]{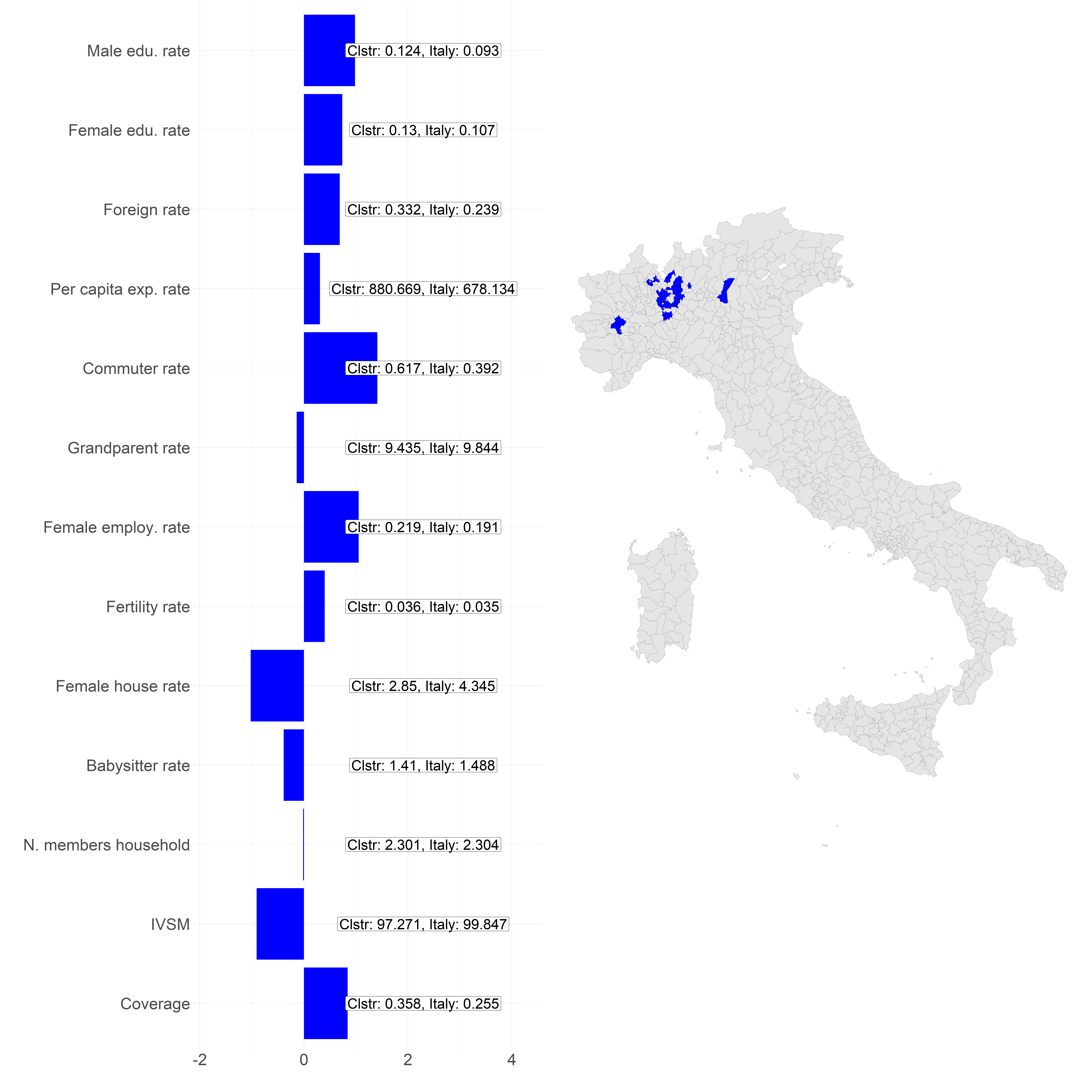}

   \end{minipage}\vfill
      \begin{minipage}{0.33\textwidth}
     \centering
     \textbf{Cluster 7 [Group 2]}\par\medskip
     \includegraphics[width=.7\linewidth]{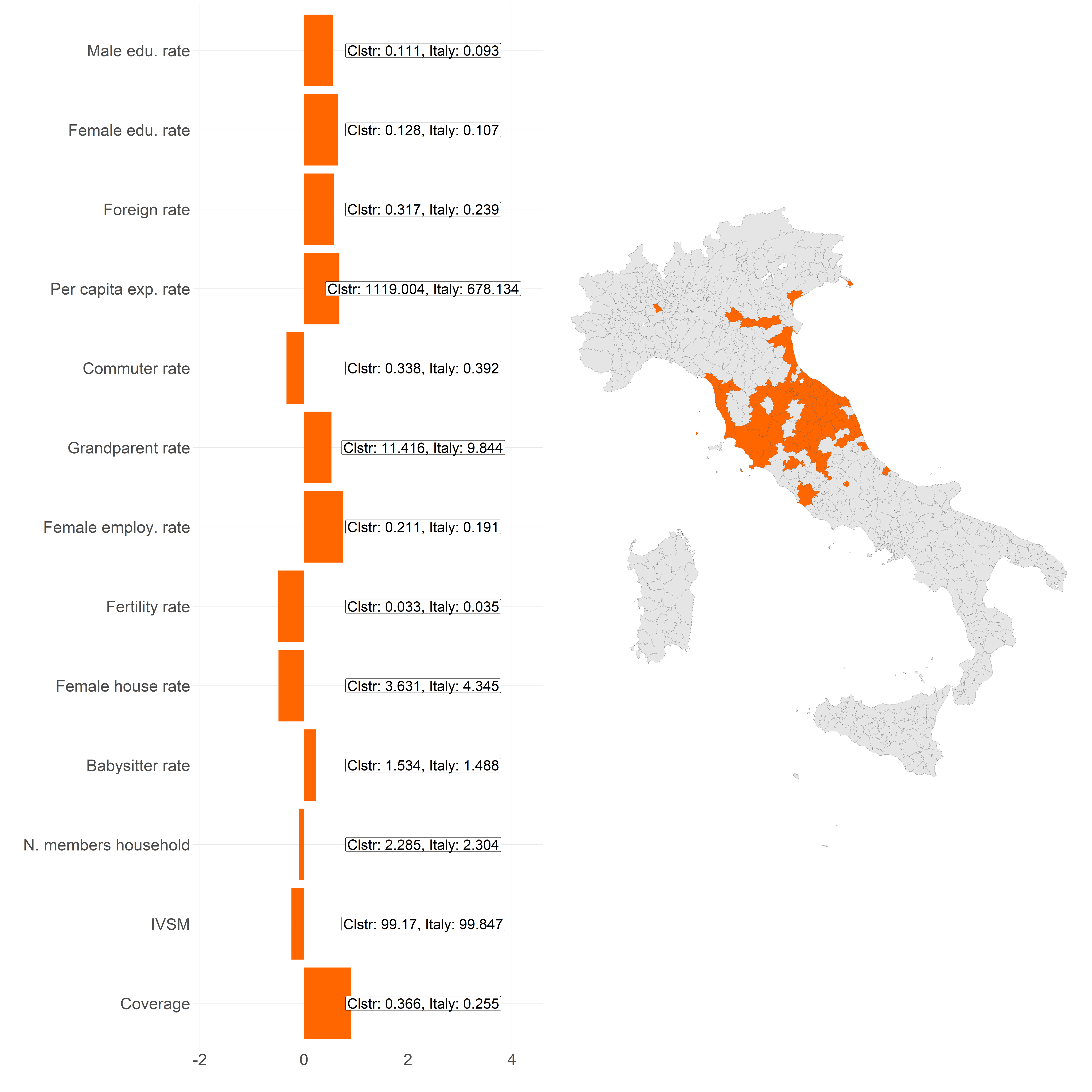}
   \end{minipage}\hfill
    \begin{minipage}{0.33\textwidth}
     \centering
     \textbf{Cluster 8 [Average cluster]}\par\medskip
     \includegraphics[width=.7\linewidth]{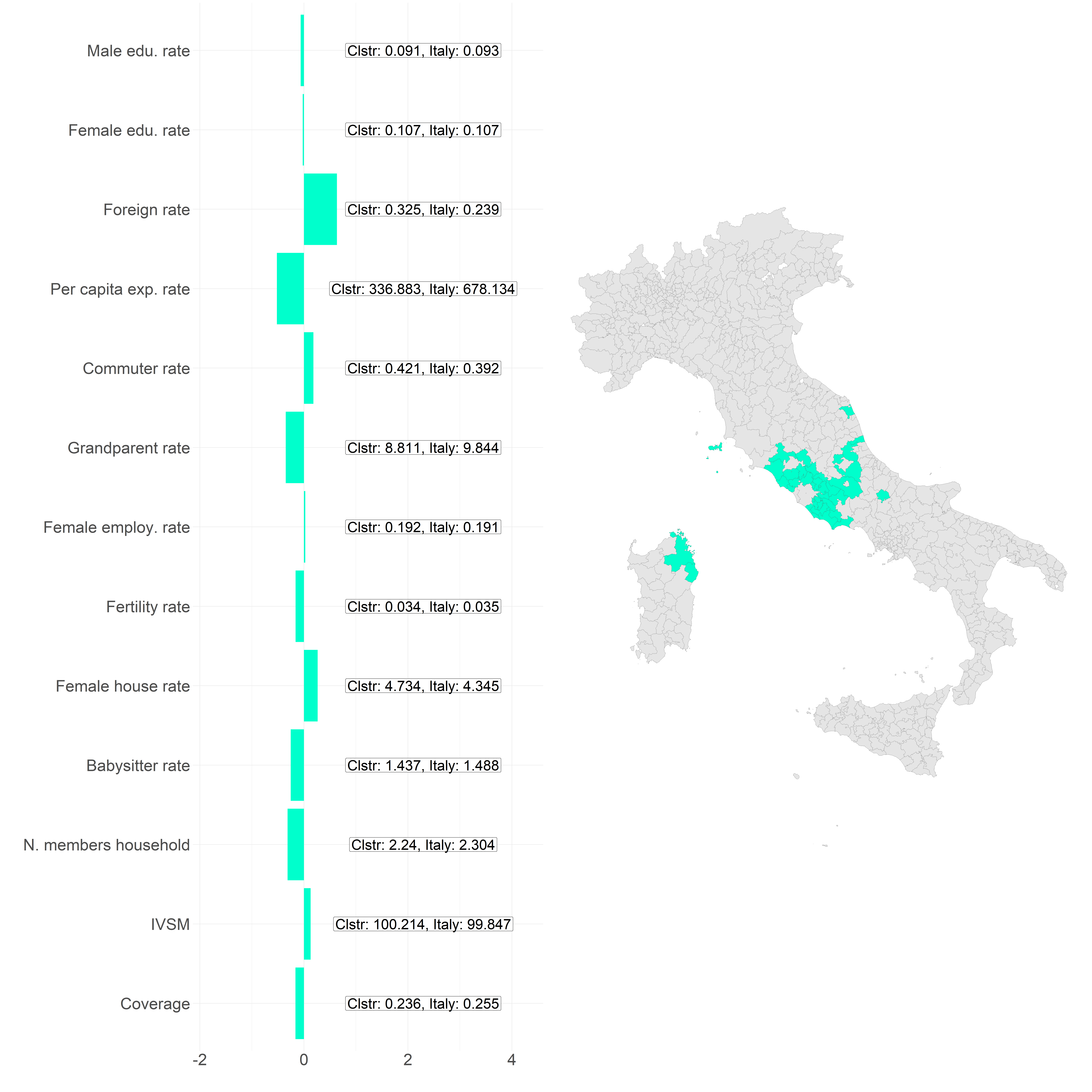}
   \end{minipage}\hfill
   \begin{minipage}{0.33\textwidth}
     \centering
     \textbf{Cluster 9 [Group 3]}\par\medskip
     \includegraphics[width=.7\linewidth]{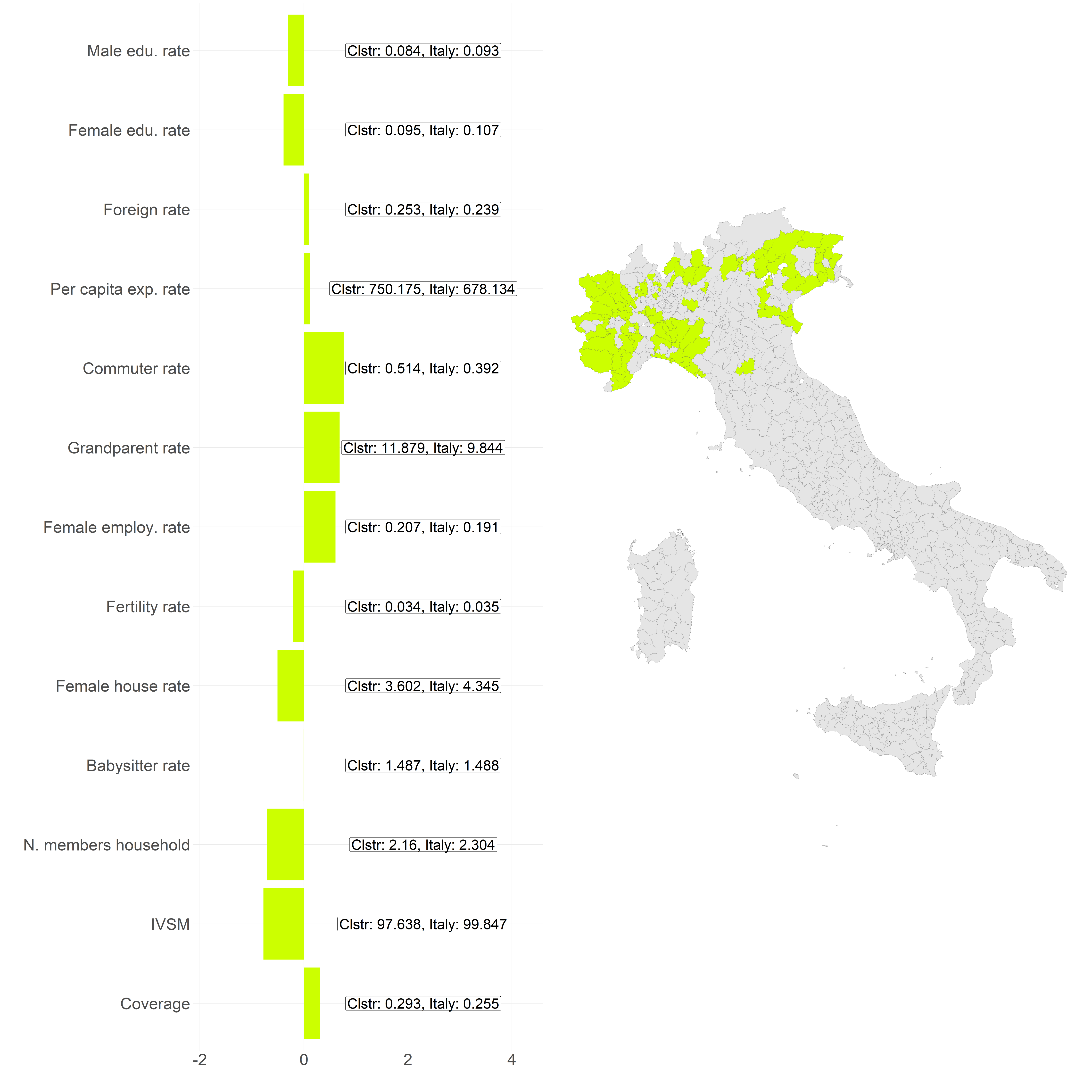}

   \end{minipage}\vfill
      \begin{minipage}{0.33\textwidth}
     \centering
     \textbf{Cluster 10 [Group 3]}\par\medskip
     \includegraphics[width=.7\linewidth]{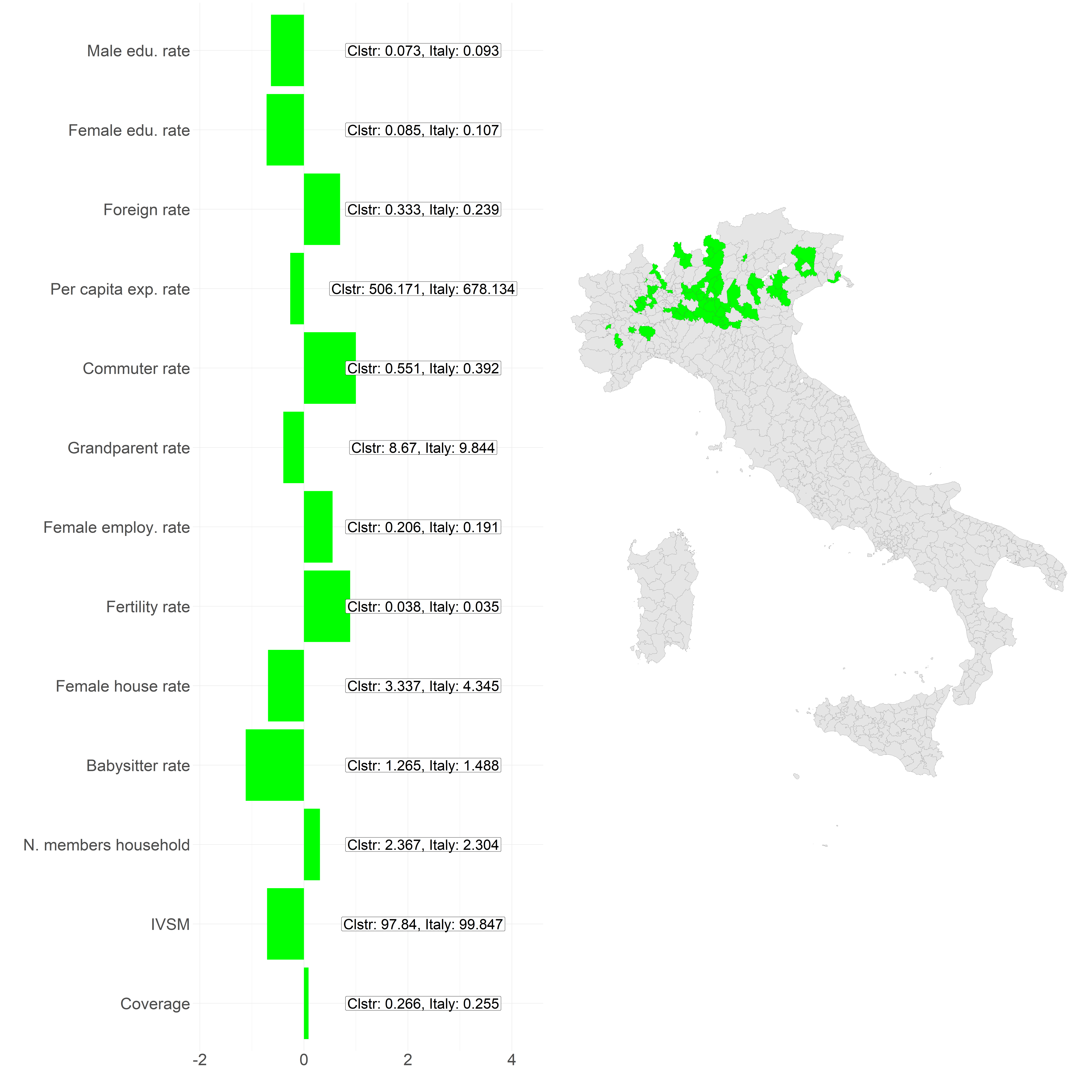}
   \end{minipage}\hfill
    \begin{minipage}{0.33\textwidth}
     \centering
     \textbf{Cluster 11 [Group 3]}\par\medskip
     \includegraphics[width=.7\linewidth]{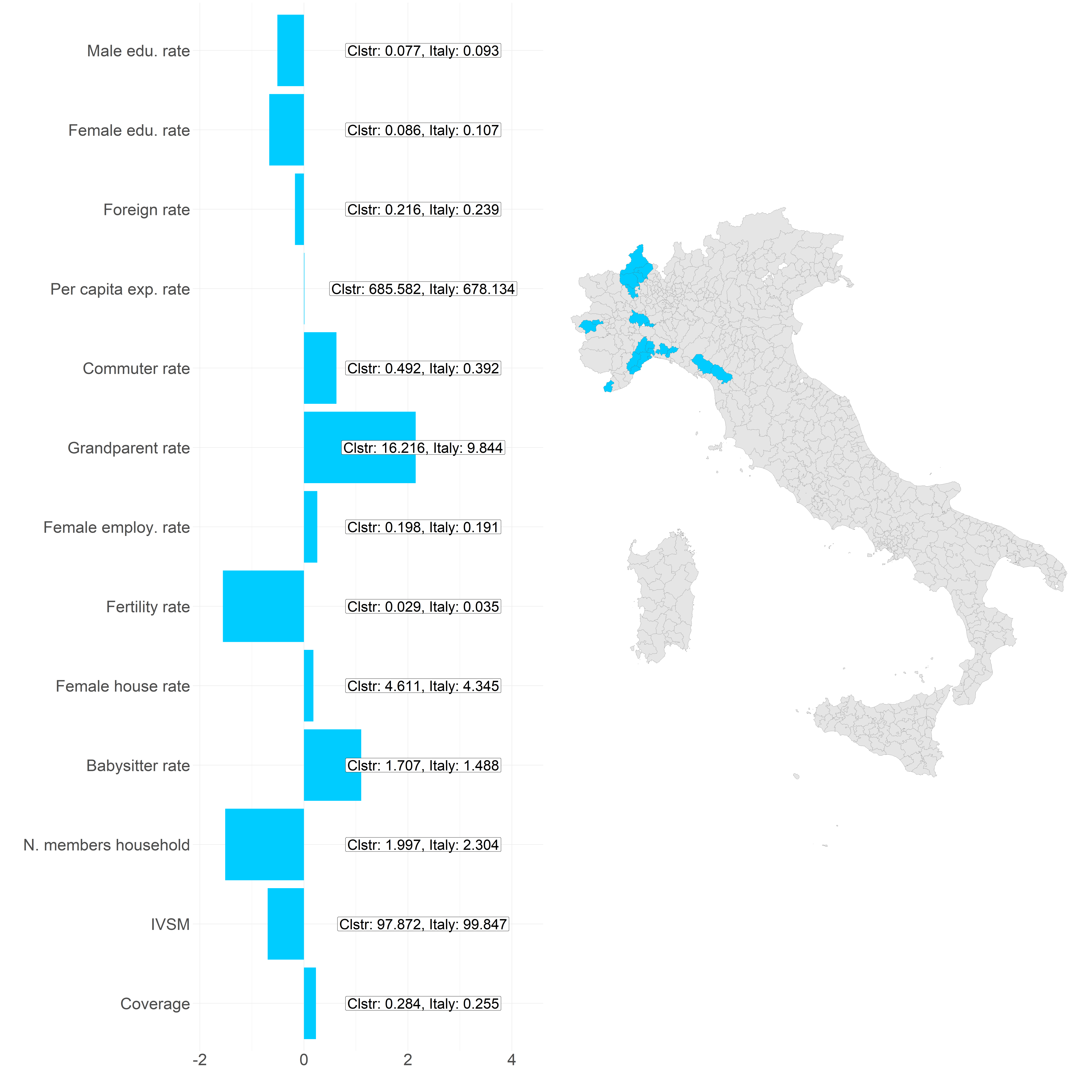}
   \end{minipage}\hfill
   \begin{minipage}{0.33\textwidth}
     \centering
     \textbf{Cluster 12}\par\medskip
     \includegraphics[width=.7\linewidth]{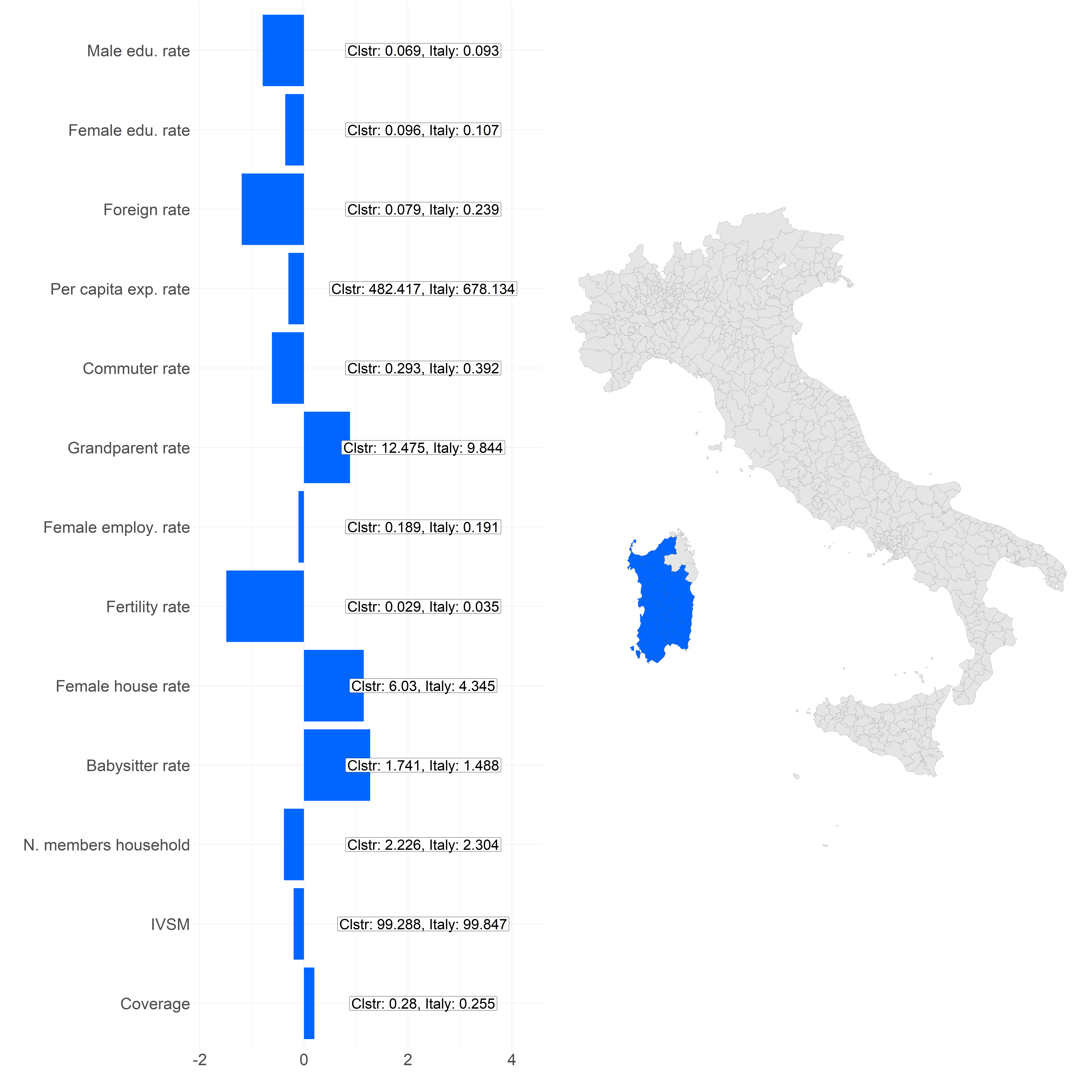}

   \end{minipage}\vfill
      \begin{minipage}{0.33\textwidth}
     \centering
     \textbf{Cluster 13}\par\medskip
     \includegraphics[width=.7\linewidth]{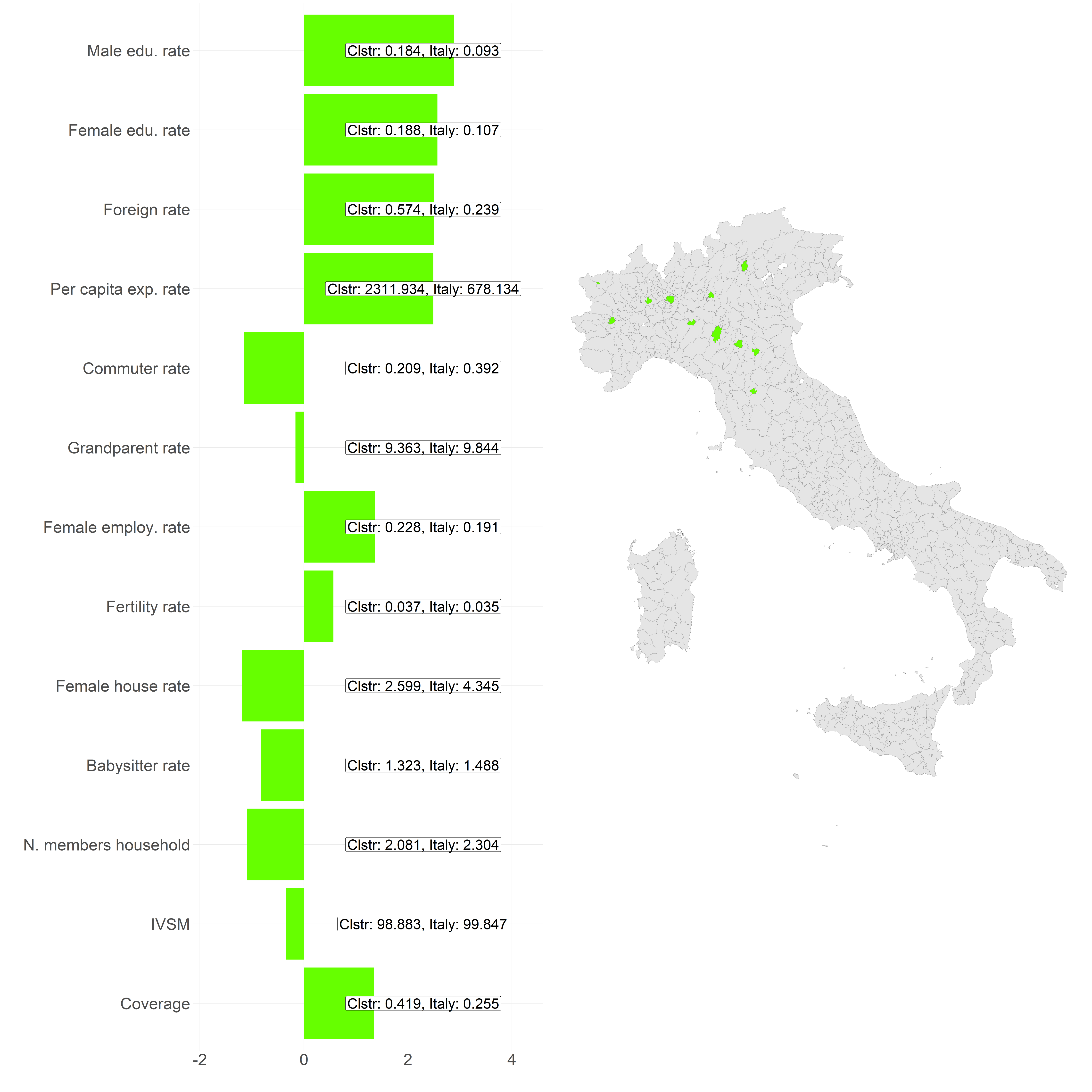}
   \end{minipage}\hfill
    \begin{minipage}{0.33\textwidth}
     \centering
     \textbf{Cluster 14}\par\medskip
     \includegraphics[width=.7\linewidth]{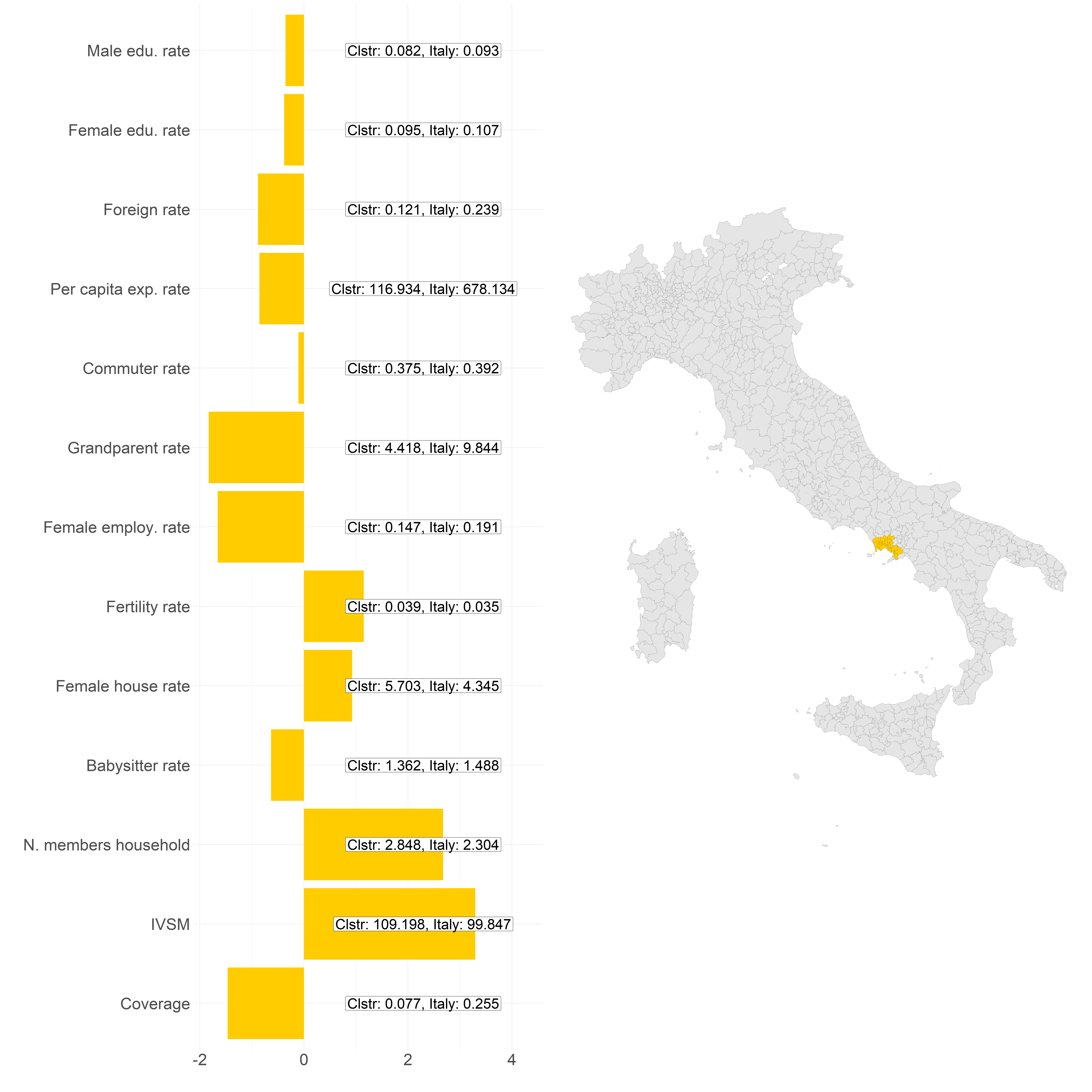}
   \end{minipage}\hfill
   \begin{minipage}{0.33\textwidth}
     \centering
     \textbf{Cluster 15}\par\medskip
     \includegraphics[width=.7\linewidth]{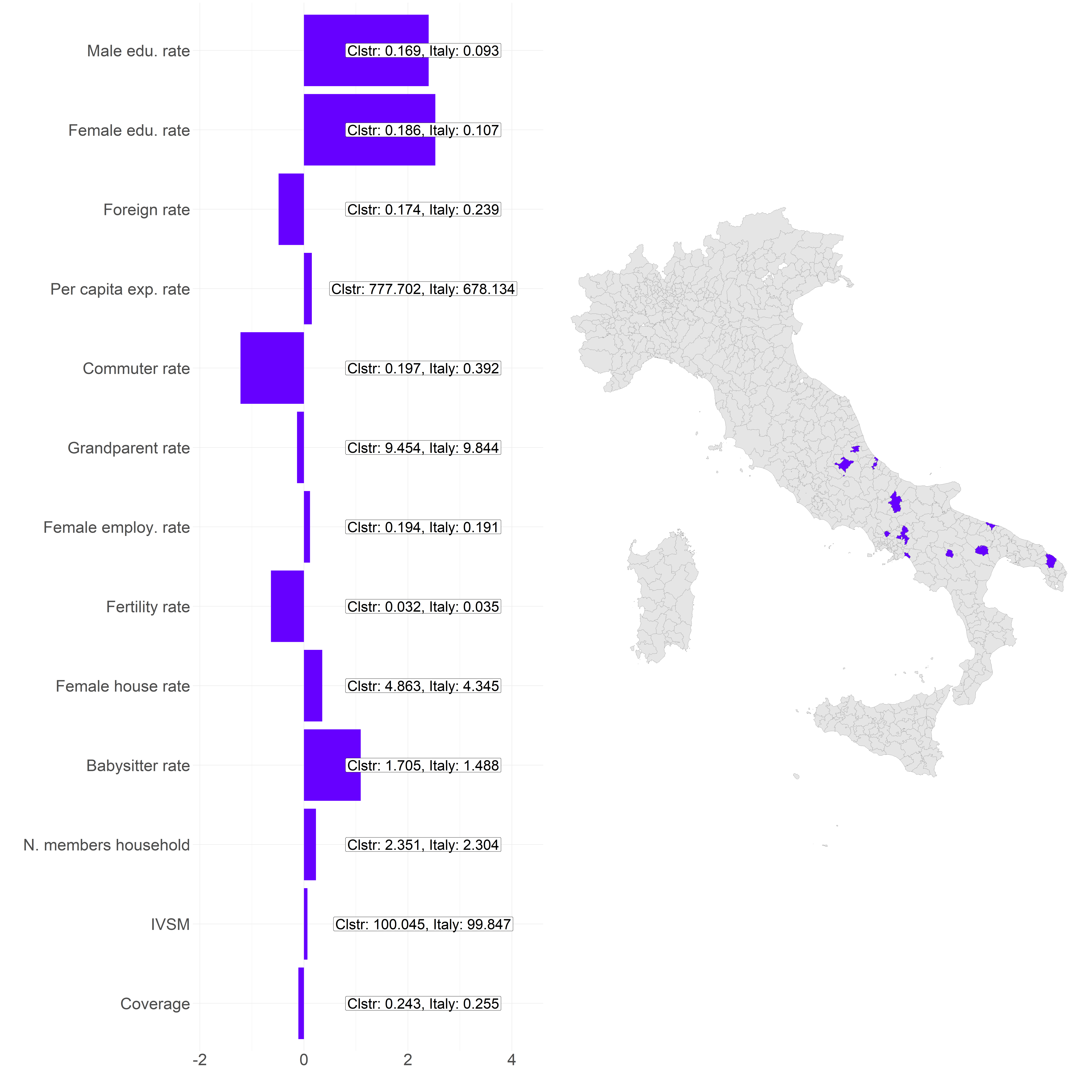}

   \end{minipage}
   \caption{Geographical representation of the $15$ clusters from the SKATER algorithm. Each plot represents by color the ATS composing the cluster. The bar plots show the mean values of the scaled variables within the clusters, while in the box, we insert the mean values of the unscaled variables within the cluster and the unscaled national averages.}\label{fig:cluster}
\end{figure}

Figure \ref{fig:cluster} describes each of these clusters. Each plot reports the ATS involved in the cluster and some information regarding the variables analyzed. The bar plots describe the averages of the scaled variables within the clusters, while the values in the box indicate the averages of the unscaled variables within the cluster and the national averages. 
Clearly, understanding the information coming from $15$ clusters is a difficult task. Still, with careful analysis, we can define four groups composed of multiple clusters and three clusters that instead have too different characteristics to be considered within one group.

The first group, composed of clusters $1$, $2$, $3$, and $4$, represents the South of Italy, characterized by low coverage, high social and material deprivation, a large number of components in the family, and a low female employment rate. However, the fourth cluster has a better childcare service situation with respect to the other three in terms of coverage, female employment rate, and social and material deprivation. This cluster is also characterized by some possible alternatives to the childcare service, i.e., babysitters and grandparents. We can also note that the first cluster includes the Sicilia region without the main municipalities, while the third one is composed of the Sicilian urbanized areas. In Sicilia, the situation of the main municipalities is worse than in suburban areas.

The second group of clusters is composed of ATS from the North and the Center (plus a part of Sardegna), i.e., the group is composed of clusters $5$, $6$, and $7$. This group of clusters is characterized by high coverage and female employment rate, as well as low social and material vulnerability index.
In addition, clusters $6$ and $7$ are also distinguished by a high level of female and male education, while cluster $5$ by a high foreign rate and a high per-capita expenditure rate. 
We can note that cluster $7$ includes the Roma municipality, while cluster $6$ includes ATS outside the main Italian municipalities like Torino and Milano, where the commuter rate is large and characterized by a lot of workers that live in the suburban areas but work in the city.

The eighth cluster can be defined as the ``average'' cluster, characterized by values of the variables near the national average.

The third group consists of clusters $9$, $10$, and $11$ and is characterized by low female and male educational qualification rates, high commuter rates, a low social and material vulnerability index, and a medium-high coverage. 
These ATS have a relatively good situation; however, cluster $11$ probably can leverage more alternatives to childcare services than the others, such as high grandparents and babysitter rates. Furthermore, these clusters can be classified as the comparatively less prosperous northern ATS, with satisfactory social and economic conditions but still inferior to the second cluster group, namely clusters $5$, $6$, and $7$. Similarly to the second group, this category encompasses ATS located outside the primary municipalities, distinguished by a significant rate of commuting.

Finally, the last four clusters should be analyzed separately. 
The cluster $12$ includes only Sardegna, characterized by an old population (i.e., low fertility rate and high grandparents rate). Instead, cluster $13$ comprises the main municipalities of North Italy, e.g., Milano and Torino. 
This cluster is characterized by a high level of education for both males and females due to the presence of important universities.
Since this cluster is composed of ATS placed in big cities, the foreign rate is also high, and the commuter rate is low, i.e., these cities are attractive to foreigners, and workers do not move to get to work. 
Finally, these ATS have a large coverage of childcare services, as well as good participation of females in the labor force. The cluster $14$ is the most deprived one and shows large values of social and vulnerability index, low coverage, and also low rates of alternative childcare services such as babysitters and grandparents. 
This cluster includes the area around the city of Napoli, also characterized by a high fertility rate and a large number of components in the household, as well as a high female home rate. Finally, the cluster $15$ is composed of ATS placed on central and southern municipalities characterized by the presence of several major universities. In fact, we can observe a high education rate for both females and males and a possibility of employment locally being the commuter rate lower than the national average. 

As a last step, we fit a penalized multinomial model \citep{glmnet} to understand the most significant variables that specify a cluster. Cluster $8$ is considered as the reference category, being the cluster that is closer to the national average. 
We report here some details on the interpretations of the model coefficients and insights into the model, while the complete coefficient estimates are reported in Table \ref{tab:multinom_glmnet} in Appendix \ref{app_results}. 

Low coverage and female employment rate strongly influence the probability of being in clusters $1$, $2$, and $3$. For example, the log relative risk ratio of being in cluster $1$ versus $8$ decreases by approximately $1.64$ for a one-unit increase in the female employment rate, and it decreases by approximately $2.04$ for a one-unit increase in coverage. The probability of being in cluster $4$ is negatively affected by the female employment rate and slightly positively affected by the coverage rate. The log relative risk with respect to cluster $8$ decreases by approximately $0.298$ for a one-unit increase in coverage. This reflects the picture that we saw in the cluster analysis: clusters $1$, $2$, $3$, and $4$ share similar characteristics but cluster $4$ has a better situation with respect to the other three clusters.
In addition, we can note a negative effect of the commuter rate in the probability of being in these clusters, with respect to cluster $8$. Even with an expanding supply, demand for childcare services in these areas might continue to be low if access is impeded by factors like high tuition fees, geographical distance, and perceived low quality of service. Therefore, policies should aim to ensure widespread accessibility of childcare services in these areas while simultaneously increasing the overall supply. This approach is essential in overcoming cultural barriers to service utilization and relieving mothers from the responsibility of early childcare, which can have a significant impact on employment rates.

Analyzing the cluster $5$, which is composed mainly of regions of the center of Italy, we can underline a strong positive effect of the per capita public expenditure rate (i.e., the log-risk ratio increases by approximately $3.49$ for a one-unit increase), and a strong negative effect of the female house rate (i.e., the log-risk ratio decreases by approximately $2.64$ for a one-unit increase). This area of the country appears to be among the most virtuous in Italy, where the historical development of public childcare services has resulted in a relatively high and evenly distributed supply in the territories and where female labor market participation is among the highest in the country. The commuter rate strongly influences the probability of being in cluster $6$ having an estimated coefficient equal to $3.836$, i.e., it is a cluster characterized by commuting workers, being, in fact, composed of areas outside the main cities of northern Italy.  

The probability of being in cluster $7$ and $9$ versus cluster $8$ increases if the female employment rate increases, as well as the grandparent and per capita expenditure rates. 
The main difference between these two clusters is the effect of the coverage and number of members in the household. 
We observe a positive effect with respect to being in cluster $8$, as the log-odd of being in cluster $7$ versus $9$ increases by $0.83$ with one-unit increases in the coverage rate. The fertility rate has a negative effect on the probability of being in both clusters $7$ and $9$. However, the relative risk ratio of being in cluster $7$ versus $9$ strongly increases if the number of members in the household increases. In these areas, childcare services must be focused on families composed of commuter working parents. One potential policy approach could involve directing childcare resources toward the municipalities where parents work, encouraging cooperation among local governments to fund educational services (or even encouraging the creation of daycare centers as part of corporate welfare). Additionally, enhancing the usability of these services is crucial. Providing flexible hours for dropping off and picking up children, as well as extending afternoon hours, can significantly boost participation in childcare services among working families. The number of members in the household also negatively affected the probability of being in cluster $11$, joined with the strong positive effect of the grandparent rates. This cluster mainly comprises ATS of the Liguria region, characterized by an aging population.

The probability of being in the region of Sardegna (i.e., cluster $12$) is negatively affected by the foreign rate. People do not travel for work compared to the national average, and a low fertility rate. The log relative risk ratio decreases by $2.5$ for one-unit increases in the fertility rate. In fact, Sardegna suffers from an accentuated process of population aging.

In these specific regions, more so than in others (although Italy as a whole is facing extremely low fertility rates), the coverage of childcare education services is witnessing a significant increase. This growth is not due to a rise in supply but rather stems from an overall decline in the number of children. To prevent the risk of under-utilization of educational services resulting from reduced user numbers and the prevalence of informal care provided by grandparents and relatives, it becomes crucial to establish conditions that encourage higher participation rates among families. This can be achieved by reducing barriers to service utilization and improving their accessibility. Additionally, alongside efforts to expand the overall supply of children's educational services, it is necessary to implement robust birth policies.

As noted in the previous section, cluster $13$ (i.e., big northern municipalities) is denoted by a negative female home rate with respect to the national mean (i.e., the log relative risk decreases by $3.24$ for a one-unit increase of the female home rate), while cluster $14$ (i.e., Napoli area) by a positive number of members in the household (i.e., the log relative risk increases by $4.24$ for a one-unit increase of the number of members in the household), and finally, cluster $15$ (i.e., universities cities) by a significantly positive effect of the male educational qualification rate (i.e., the log relative risk increases by $2.66$ for a one-unit increase of the number of members in the household) and negative effect of the commuter rate (i.e., the log relative risk decreases by $2.13$ for a one-unit increase of the number of members in the household). Policymakers can take these characteristics into account to address design welfare policies, i.e., cluster $15$ could implement policies aimed at promoting the birth rate among young couples who are attending universities. This can be accomplished by facilitating residency choices and providing childcare education services not only for local students and graduates but also for those coming from outside the region.

\begin{figure}[!htb]
     \centering
     \includegraphics[width=\linewidth]{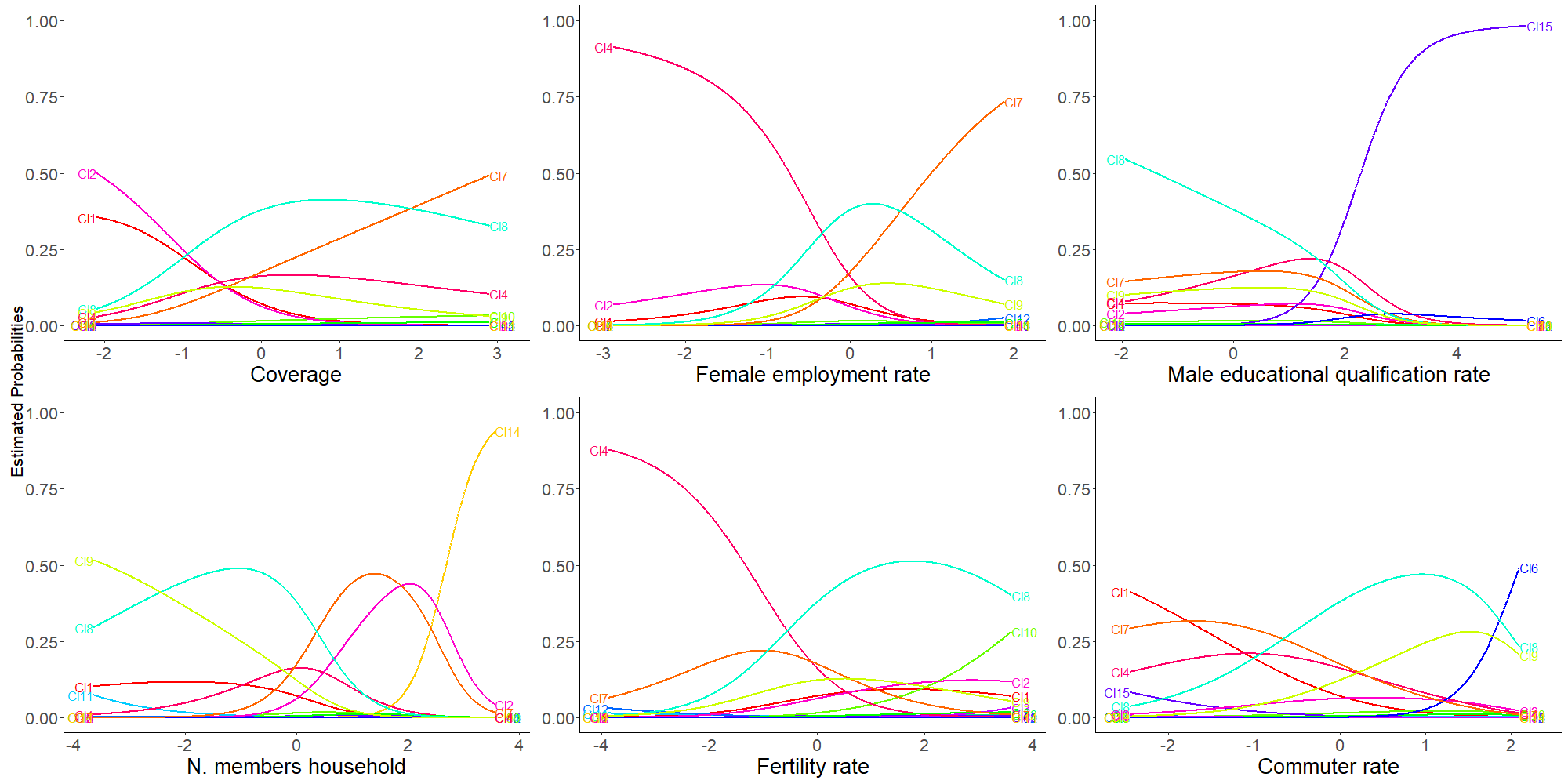}

   \caption{Predicted probabilities of being in clusters $j$, where $j = 1, \dots, 15$ across several values of 
 coverage (left top figure), female employment rate (center top figure), male educational qualification rate (right top figure), number of members in the household (left bottom figure), fertility rate (center bottom figure) and commuter rate (right bottom figure).}\label{fig:multinom}
\end{figure}

To extend these conclusions, panels in Figure~\ref{fig:multinom} show the predicted probabilities of being in one cluster across different values of some of the covariates inserted into the model. The remaining plots are reported in Appendix \ref{app_results}.

Considering a scaled coverage equals $3$, the probability of being in cluster $7$ equals approximately $0.5$ while having a high negative coverage leads to a probability of being in clusters $1$ and $2$ equals at least $0.4$. Instead, looking at the top center of Figure \ref{fig:multinom}, we can note how the probability of being in cluster $7$ increases if the female employment rate increases and how this probability decreases if we consider cluster $4$ instead. It is clear now that cluster $15$ is characterized by a high percentage of males with high degrees. The probability of being in this cluster equals $1$ for high male educational qualification rate values.

Looking instead at the left bottom of Figure \ref{fig:multinom}, there is clear evidence of the covariates defining cluster $14$. The families in the Naples area are composed of many individuals. The probability of being in this cluster equals $1$ when the number of household members is large. 
On the other hand, the opposite situation can be observed when analyzing cluster $9$. Low values of a commuter rate increase the probability of being in cluster $1$, while high values increase the probability of being in cluster $6$. Finally, the probability of being in cluster $4$ increases with low values of the fertility rate, while the probability of being in cluster $10$ increases with high values of fertility rate.

These analyses clearly reveal the variability in the combination of supply and demand for childcare services in Italy. Drawing on the analyses presented here, it becomes possible to design public policies that are specifically suited to the distinct characteristics of each territory. These policies should go beyond considering the mere presence of childcare educational service provision and take into account the socio-demographic and economic characteristics of the local population. These factors play a crucial role in shaping public policies that successfully promote and increase family participation rates in these services.

\section{Conclusions}\label{conclusions}

Considering the entire Italian territory, the average coverage rate stands at $27.2$ places for every $100$ toddler. 
Such a level of coverage is then still below the level of supply indicated as the European minimum target and taken up by the Italian National Recovery and Resilience Plan (NRRP) itself, fixed at $33$ authorized places for every $100$ child.
Disaggregating the national average figures, strong differences emerge between territorial divisions, with large geographical areas such as North-East and Central Italy presenting on average coverage above the European target, the North-West close to the target, while the South and the Islands are still far from the target. 
More in-depth analysis has shown that differences in supply levels are present not only between macro areas and regions but also within them, e.g., peripheral areas versus urban centers and mountainous areas versus non-mountainous ones. 
This variability seems to derive from historical factors, both related to demand dynamics (e.g., falling birth rate, female labor participation) and deriving from policies of local and regional governments in equipping themselves with this type of social/educational supply due to the absence of a national governance (and in many regions of regional governance) effective in influencing its widespread and homogeneous development. 
Only in a few regions is there a spatially uniform distribution of services; in most of them (including several regions in the North), the supply of services is concentrated in some territories more than in others, and the latter, in some cases, are completely lacking. Moreover, Italy is facing ultra-low fertility rates. In this context, the Italian government has approved an extraordinary funding plan, utilizing Next Generation EU funds to bolster the provision of early childhood education services. The aim is to address the imbalances in supply levels across different regions of the country.

In this challenging environment, this paper has explored the spatial distribution of the childcare service in Italy, jointly with territorial socio-demographic information. Thanks to the application of the spatial clustering approach, we were able to show and explore the strong variability of demand and supply across the Italian territory. We noted how the situation in southern Italy is critical: the low services provision goes together with significant material and social deprivation with few job opportunities for females that often match with a higher provision of childcare services. In particular, only in some areas, there are possible alternatives to missed educational services as a family or external support (e.g., babysitters). In contrast, northern and central Italy has more employment opportunities for females. In particular, several areas in central Italy are inhabited by large families, thus, suggesting the presence of family support, while commuting workers characterize several areas in the north. In the latter case, specific policies should be implemented to help parents organize a working life outside their municipality of residence. Finally, areas with particular characteristics were observed, such as university cities where the rate of study is high and commuting is relatively uncommon, the Naples area with considerable social and material deprivation, and suburban areas characterized by high commuting. Policymakers must observe and analyze these features to make efficient new welfare policies.

Current empirical findings pointed out how the organization of children's educational services in Italy is very different, evidently as a result of a lack of governance at the national and, often, even regional levels, which is essential for several reasons. 
Well-organized governance could favor nursery school policies and practices to be coordinated, ensuring consistency with regional goals and priorities regarding the education and care of children. 
Well-thought governance can also help to ensure that preschools are subject to uniform minimum standards and requirements regarding quality, safety, and child welfare. In addition, funding could be managed and coordinated to guarantee that they are used efficiently and fairly; the services would be more accessible to all children regardless of their socio-economic or geographical situation. 

In summary, clear governance for childcare services is important to coordinate, standardize, finance, make accessible, and support kindergartens to provide high-quality education and care for children. A well thought and tailored governance must be based on clear information about the different supply and demand of services and, more importantly, the factors that influence them. For these reasons, it is important to understand and manage the large variability of supply and demand of childcare services within and among the regions. This work highlights a need for governance federalism that policymakers must consider when new welfare policies are being planned.
For example, the NRRP should consider the complex and multidimensional structure of the supply and demand of childcare services instead of focusing only on increasing the supply. Socio-economical aspects of the population clearly characterize the demand, and the NRRP should therefore aim to create homogeneous models of supply in a wider area than the municipality.

Another contribution of our work is to show how some sophisticated statistical methods could help to offer valuable information even when data are limited and can come from different sources. Nevertheless, some limits stand. Clearly, describing childcare demand is challenging in Italy since the Italian permanent census is the only available data at the moment. 
In this paper, an attempt was made to capture the presence of an alternative to childcare through the construction of indices such as the grandparent rate and the babysitter rate. 
However, the grandparent rate includes a portion of the population that is too old to be helpful to the family, and the babysitter rate considers only female students between the ages of $15$ and $25$. 
A future direction could be, therefore, improving the description of the presence of alternative childcare if additional data become accessible. Finally, other measures could be inserted into the regionalization model to improve the description of the socio-economic structure of the population, such as the Italian citizenship income and fertility rate trends that were not available when the analysis of this manuscript was made.

\section*{Acknowledgments}

Angela Andreella gratefully acknowledges funding from the grant PON 2014-2020/DM 1062 of the Ca’ Foscari University of Venice, Italy. Authors acknowledge the agreement signed among the University Ca' Foscari of Venice, the National Institute of Statistics, and the Department of Family Policies at the Presidency of the Ministerial Council that supported and partially funded their work.  

\section*{Authors Contributions}
\textbf{Angela Andreella}: conceptualization, software, data curation, formal analysis, investigation, and writing - original draft, and writing - review \& editing. \textbf{Emanuele Aliverti}: conceptualization, investigation, writing - review \& editin, \textbf{Federico Caldura}: conceptualization, writing - review \& editing, \textbf{Stefano Campostrini}: conceptualization, writing - review \& editing and supervision.

\section*{Declaration of Competing Interest}
The authors declare no competing interests.

\section*{Additional information}
All the figures of the manuscript except Figure \ref{fig:sil} must be printed in color.

\clearpage

\bibliography{bibliography}
\bibliographystyle{apalike}

\clearpage

\appendix

\section{Exploratory plots}\label{app}

\begin{figure}
    \centering
    \includegraphics[width=\linewidth]{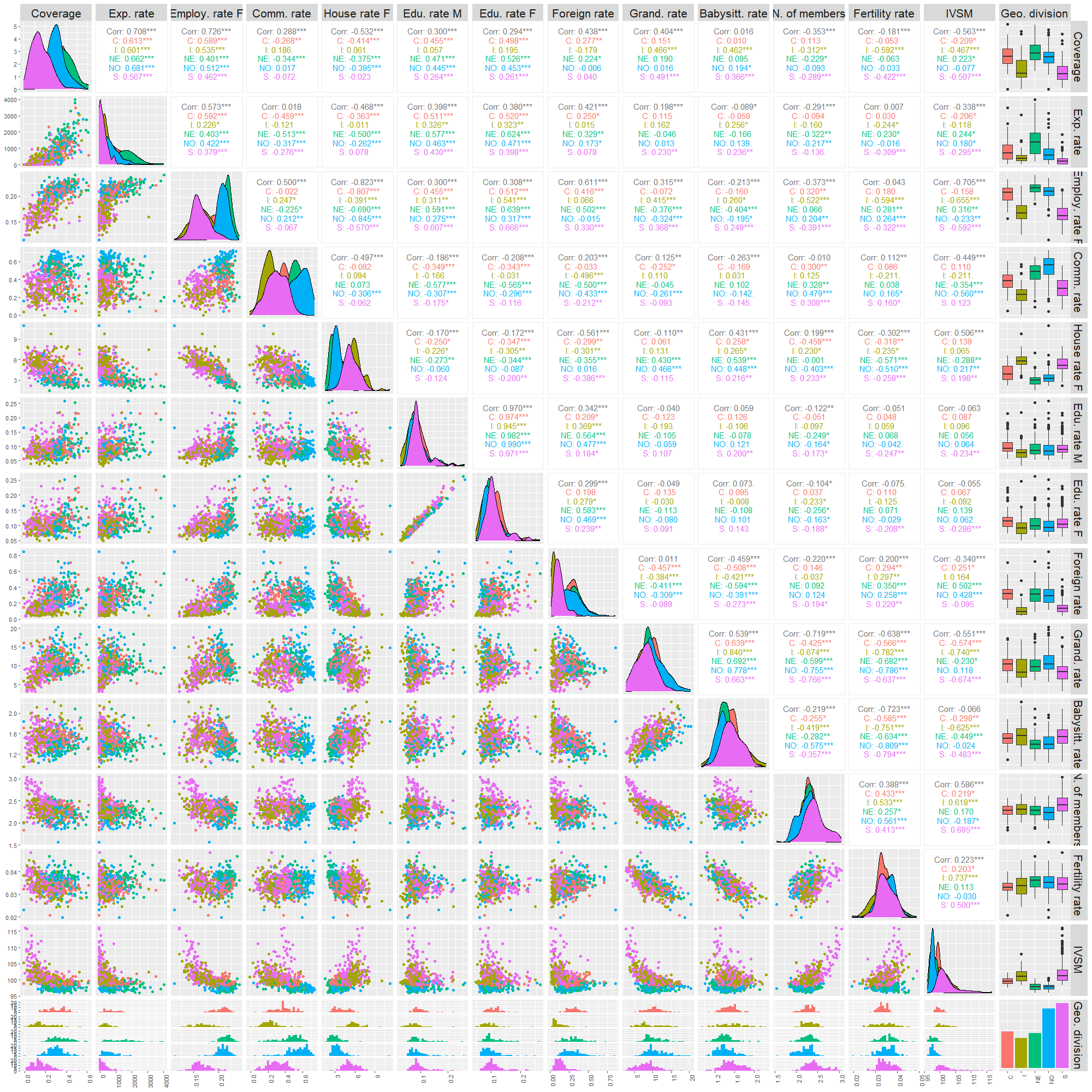}
    \caption{Exploratory plots (i.e., scatter plots, histograms, and boxplots) for each variable described in Table \ref{data} divided by macro areas (Center, Islands, North-East, North-West, and South.).}
    \label{fig:full_plot}
\end{figure}

Figure \ref{fig:full_plot} shows some exploratory plots (i.e., pairwise scatter plots, frequency histograms, and boxplots) for examining the variables defined in Table \ref{data} and used in the clustering/modeling analysis. Each color represents one Italian macro area (light red for Center, dark yellow for Islands, green for North-East, light blue for North-West, and purple for South).

\begin{figure}[!htb]
   \begin{minipage}{0.45\textwidth}
     \centering
     \includegraphics[width=\linewidth]{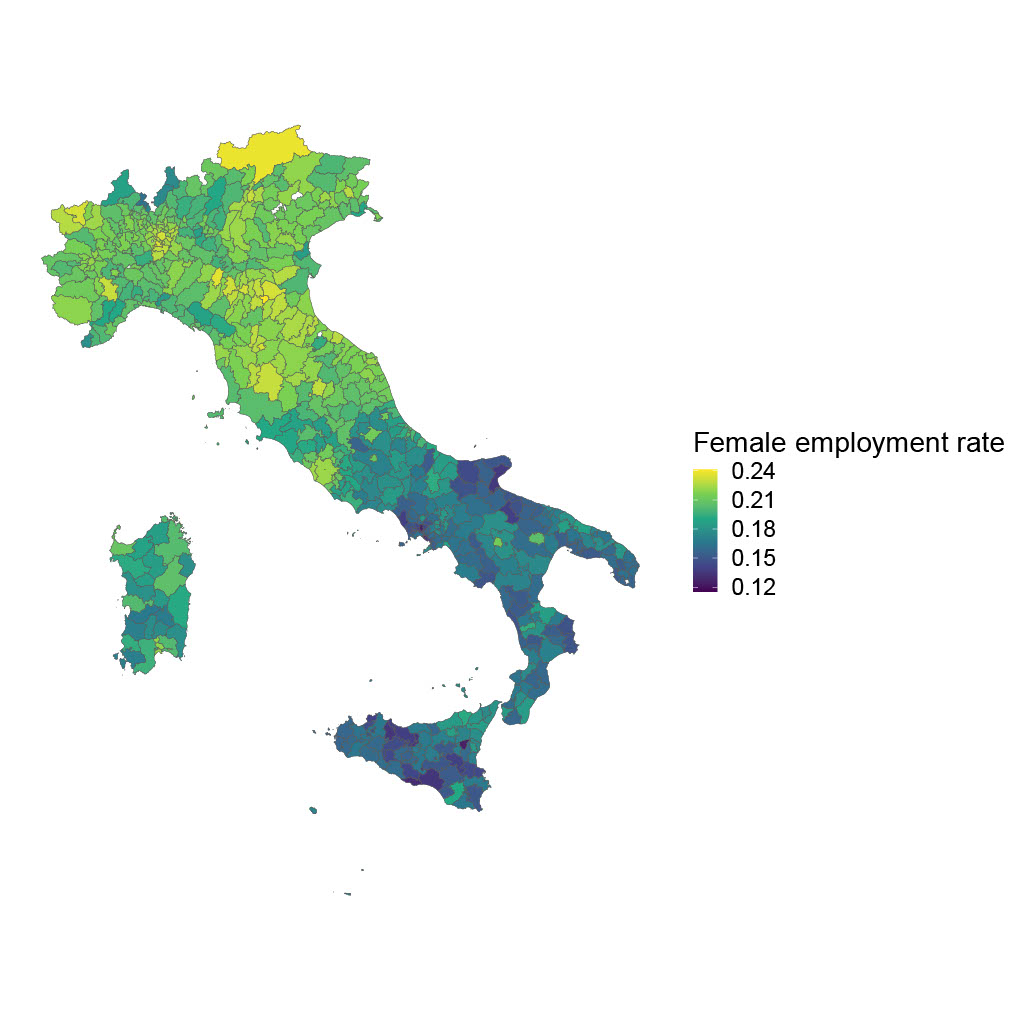}
   \end{minipage}\hfill
    \begin{minipage}{0.45\textwidth}
     \centering
     \includegraphics[width=\linewidth]{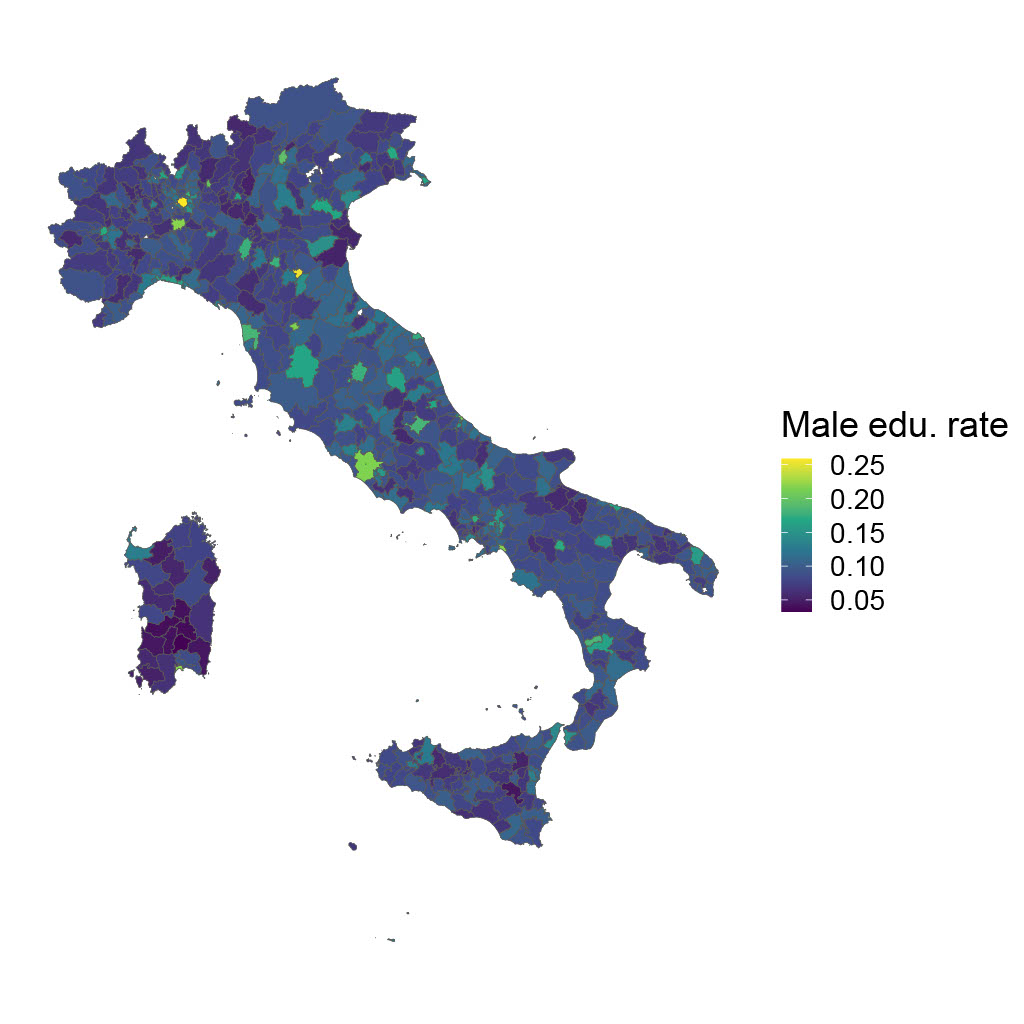}
   \end{minipage}\vfill
    \begin{minipage}{0.45\textwidth}
     \centering
     \includegraphics[width=\linewidth]{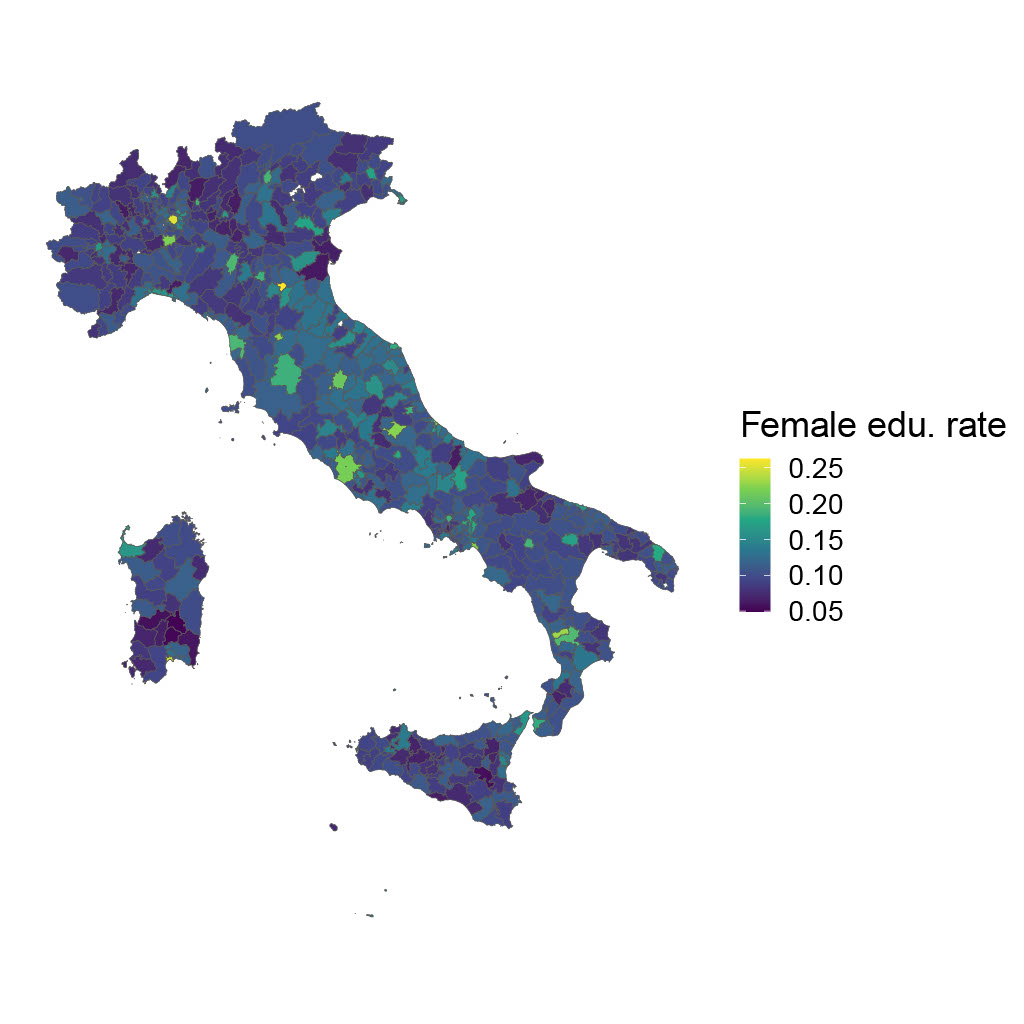}
   \end{minipage}\hfill
    \begin{minipage}{0.45\textwidth}
     \centering
     \includegraphics[width=\linewidth]{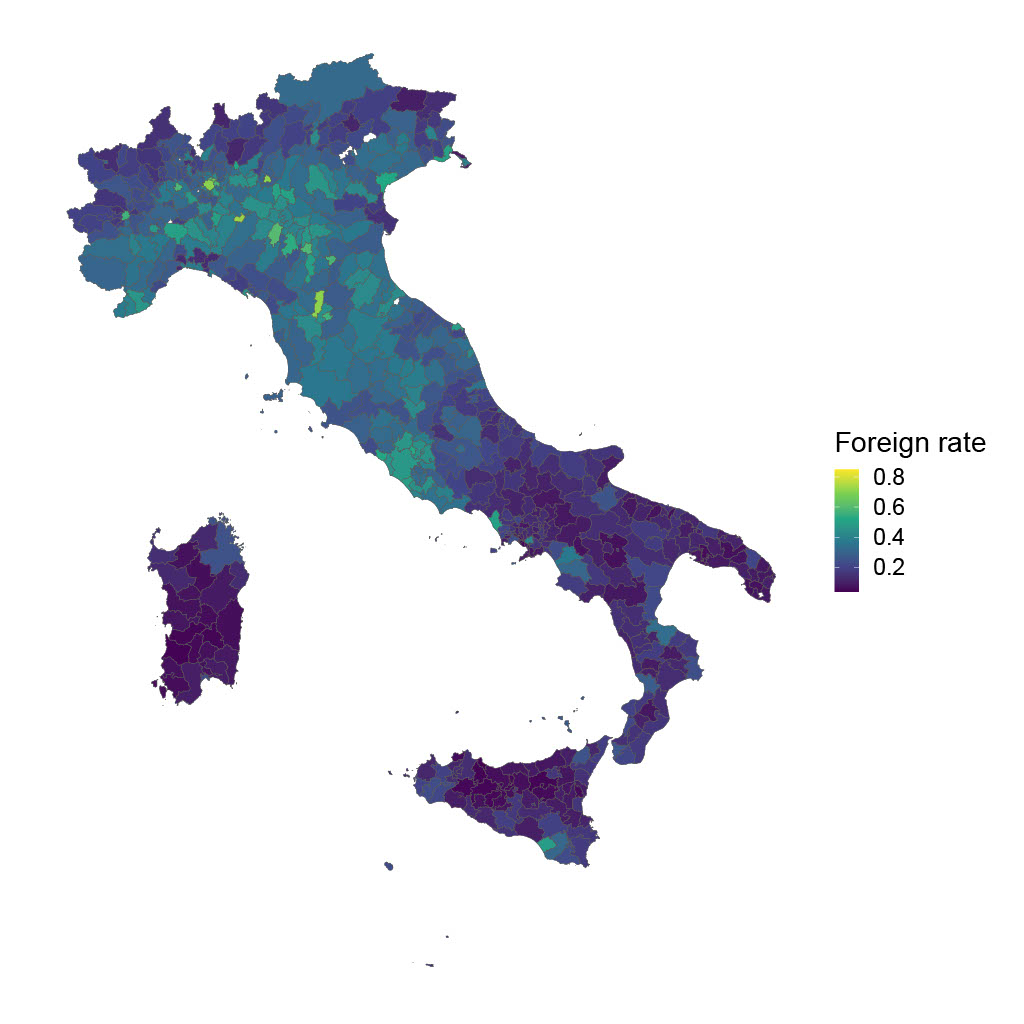}
   \end{minipage}
      \caption{Geographical map representation of the female employment rate (left top figure), male educational rate (right top figure), female educational rate (left bottom figure), and foreign rate (right bottom figure) at ATS spatial level.}\label{fig:supp1}
      \end{figure}

\begin{figure}[!htb]
    \begin{minipage}{0.45\textwidth}
     \centering
     \includegraphics[width=\linewidth]{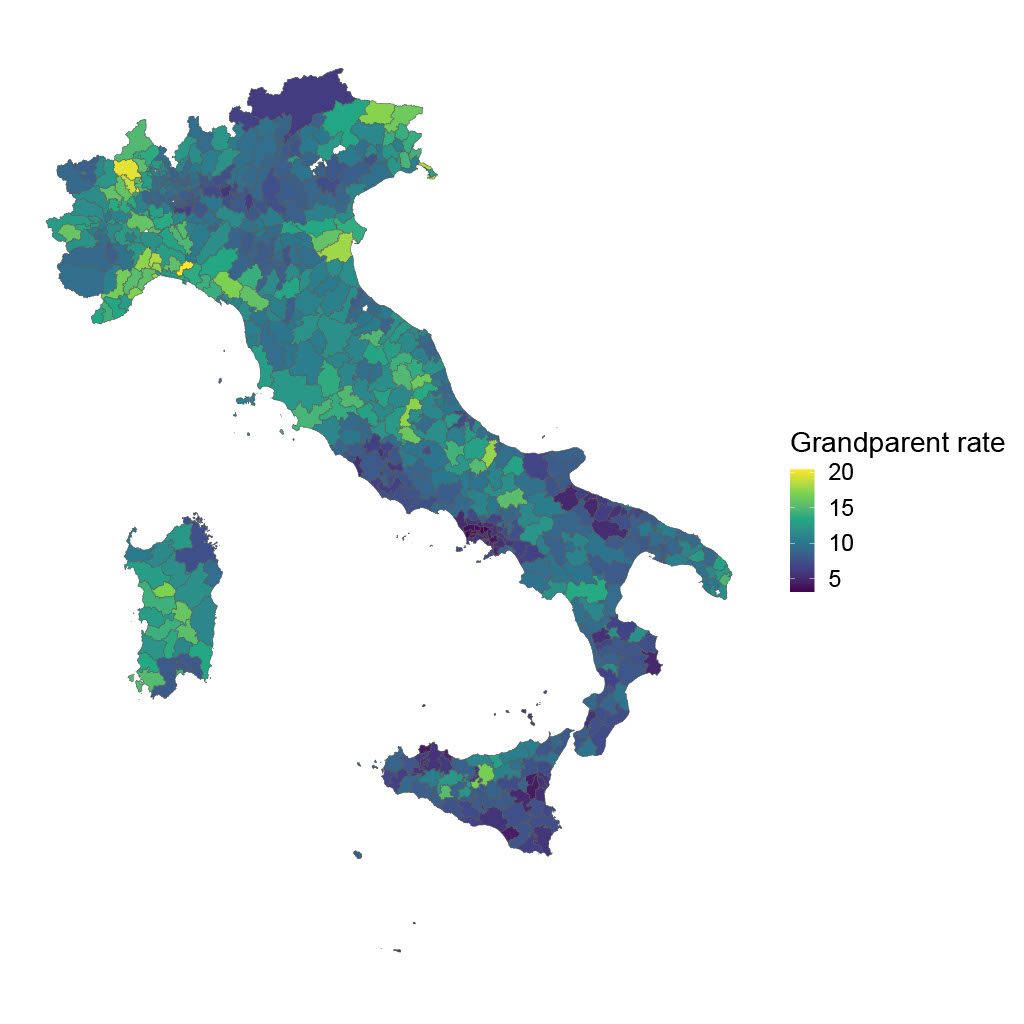}
   \end{minipage}\hfill
    \begin{minipage}{0.45\textwidth}
     \centering
     \includegraphics[width=\linewidth]{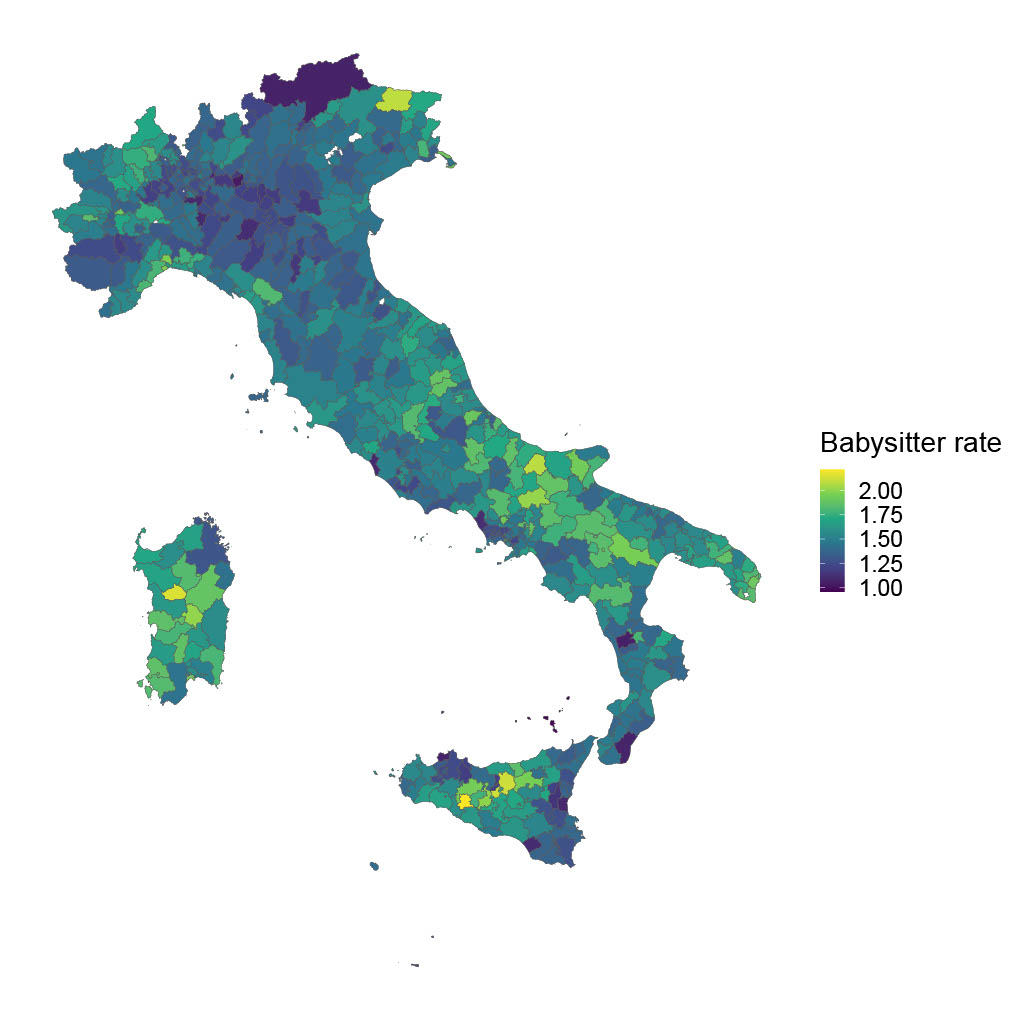}
   \end{minipage}\vfill
    \begin{minipage}{0.45\textwidth}
     \centering
     \includegraphics[width=\linewidth]{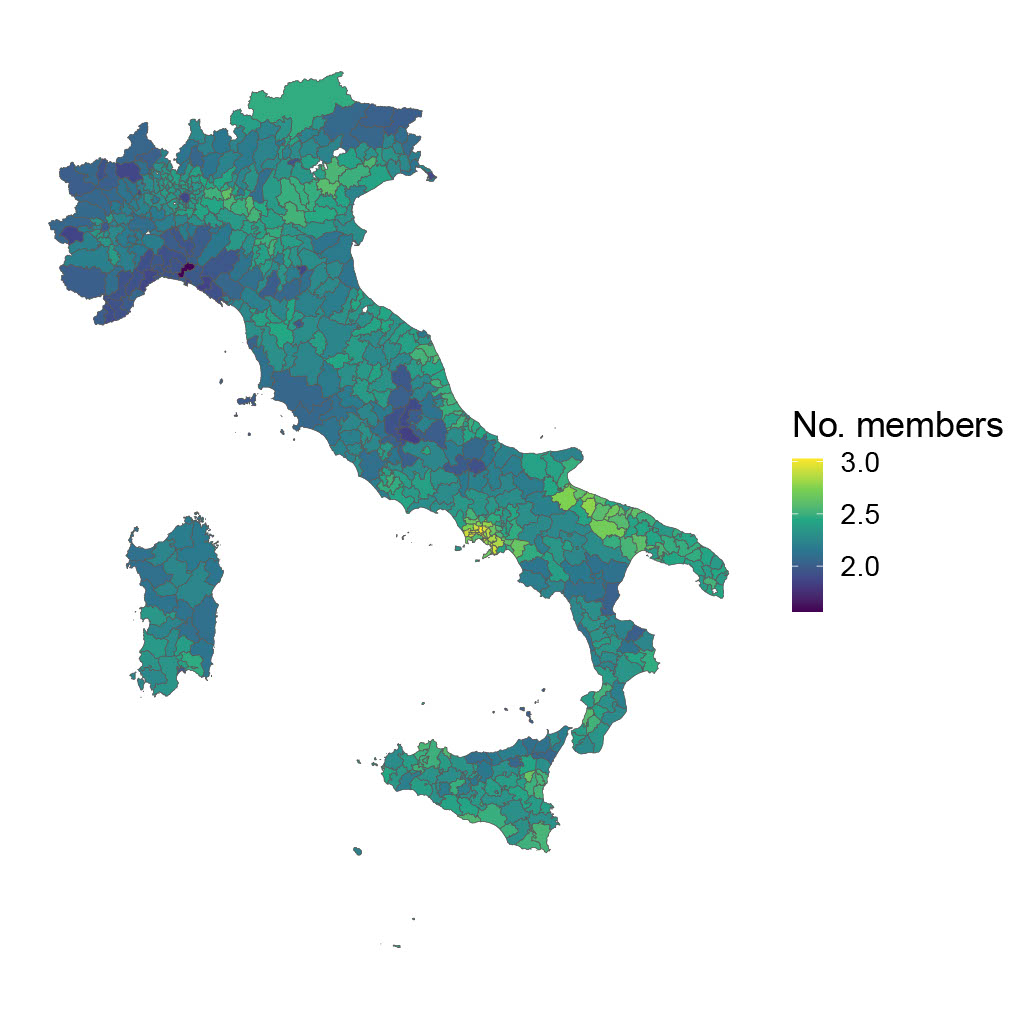}
   \end{minipage}\hfill
    \begin{minipage}{0.45\textwidth}
     \centering
     \includegraphics[width=\linewidth]{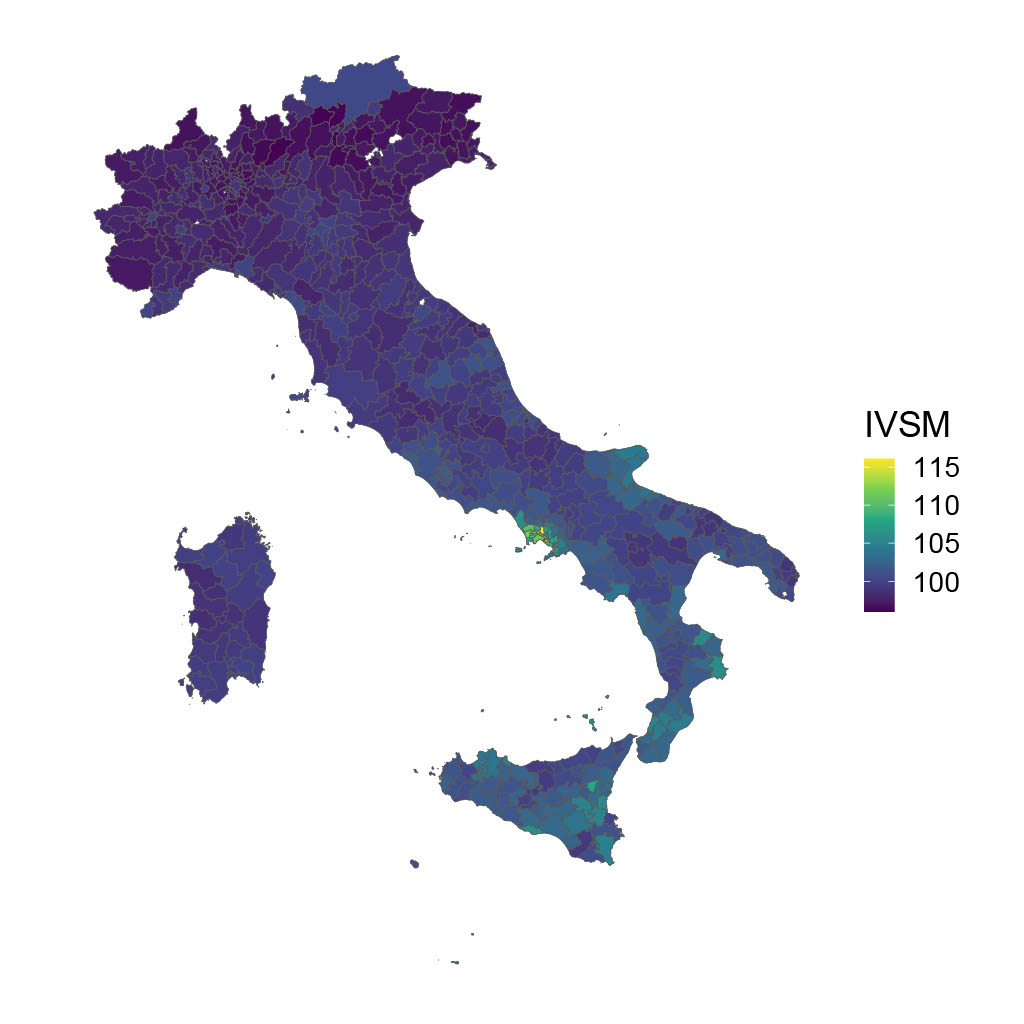}
   \end{minipage}
   \caption{Geographical map representation of the grandparents rate (left top figure), babysitter rate (right top figure), number of members in the household (left bottom figure), and IVSM (right bottom figure) at ATS spatial level.}\label{fig:supp2}
\end{figure}

Figures \ref{fig:supp1} and \ref{fig:supp2} show the geographical distribution at ATS spatial level of some variables described in Table \ref{data}. 

\section{Results}\label{app_results}

\begin{table}
\centering
\resizebox{\linewidth}{!}{
\begin{tabular}{lrrrrrrrrrrrrrrr}
\toprule
  & Cluster 1 & Cluster 2 & Cluster 3 & Cluster 4 & Cluster 5 & Cluster 6 & Cluster 7 & Cluster 8 & Cluster 9 & Cluster 10 & Cluster 11 & Cluster 12 & Cluster 13 & Cluster 14 & Cluster 15\\
\midrule
Intercept & 2.307 & 2.181 & -3.680 & 3.139 & -0.455 & -2.135 & 3.218 & 3.989 & 2.858 & 0.740 & -1.603 & -2.104 & -2.877 & -5.159 & -0.420\\
Coverage & -1.262 & -1.485 & -0.389 & 0.322 & 0.795 & 0.123 & 0.833 & 0.427 & 0.000 & 0.754 & -0.052 & 0.000 & 0.000 & -0.137 & -0.512\\
Per capita expediture rate & 0.000 & -0.874 & 0.000 & -1.127 & 1.823 & 0.000 & 0.333 & -1.667 & 0.118 & -0.088 & 0.000 & -0.426 & 1.079 & 0.000 & 0.000\\
Female employment rate & -1.369 & -1.961 & -2.770 & -2.529 & 0.889 & 0.000 & 1.254 & 0.000 & 0.197 & -0.176 & 0.167 & 2.289 & 0.000 & 0.000 & 0.000\\
Female house rate & 0.000 & 0.172 & 1.171 & -0.753 & -1.740 & -1.412 & -0.161 & 0.901 & -0.119 & 0.854 & 0.229 & 1.202 & -2.336 & 0.000 & 0.000\\
\addlinespace
Commuter rate & -1.120 & 0.305 & -1.829 & -0.366 & 1.031 & 3.836 & -0.606 & 0.556 & 1.038 & 0.711 & 1.025 & -1.083 & 0.000 & 0.000 & -1.578\\
Male educational qualification rate & -0.117 & 0.135 & 0.000 & 0.274 & -0.076 & 1.914 & 0.000 & -0.288 & 0.000 & 0.000 & 0.000 & -0.354 & 1.257 & 0.000 & 2.371\\
Female educational qualification rate & 0.722 & 0.000 & -0.071 & 0.000 & 0.000 & 0.000 & 0.512 & -0.145 & -0.485 & -1.866 & -0.931 & 0.393 & 0.000 & 0.000 & 0.000\\
Foreign rate & -1.844 & -0.315 & -1.209 & -1.422 & 1.047 & 0.891 & 0.213 & 0.474 & 0.000 & 0.628 & 0.000 & -5.606 & 1.907 & 0.000 & -1.907\\
Grandparent rate & 0.000 & -0.242 & -0.756 & 0.109 & 1.832 & 0.000 & 1.345 & -0.448 & 0.390 & 0.000 & 1.414 & 0.696 & 0.000 & -0.287 & 0.000\\
\addlinespace
Babysitter rate & 0.439 & 0.000 & -1.309 & 0.686 & -1.077 & 0.000 & 0.000 & -0.435 & 0.000 & -2.630 & 0.490 & 0.000 & 0.000 & 0.000 & 1.257\\
Number of members in the household & -0.864 & 2.001 & 0.450 & 0.000 & 1.335 & -0.521 & 1.448 & -0.693 & -1.157 & 0.518 & -1.848 & -0.880 & 0.000 & 4.929 & 0.876\\
Fertility rate & -0.018 & 0.160 & 1.435 & -1.532 & 0.357 & 0.000 & -0.838 & 0.000 & -0.245 & 0.804 & -1.691 & -2.021 & 0.922 & 0.367 & 0.000\\
IVSM & 2.567 & 0.000 & 1.066 & 0.000 & -0.242 & -2.200 & 0.000 & 1.083 & -3.103 & -3.097 & -1.841 & 0.000 & 0.000 & 3.949 & 0.000\\
\bottomrule
\end{tabular}}
\caption{Estimated coefficients from penalized multinomial model.}\label{tab:multinom_glmnet}
\end{table}

\begin{figure}[!htb]
     \centering
     \includegraphics[width=\linewidth]{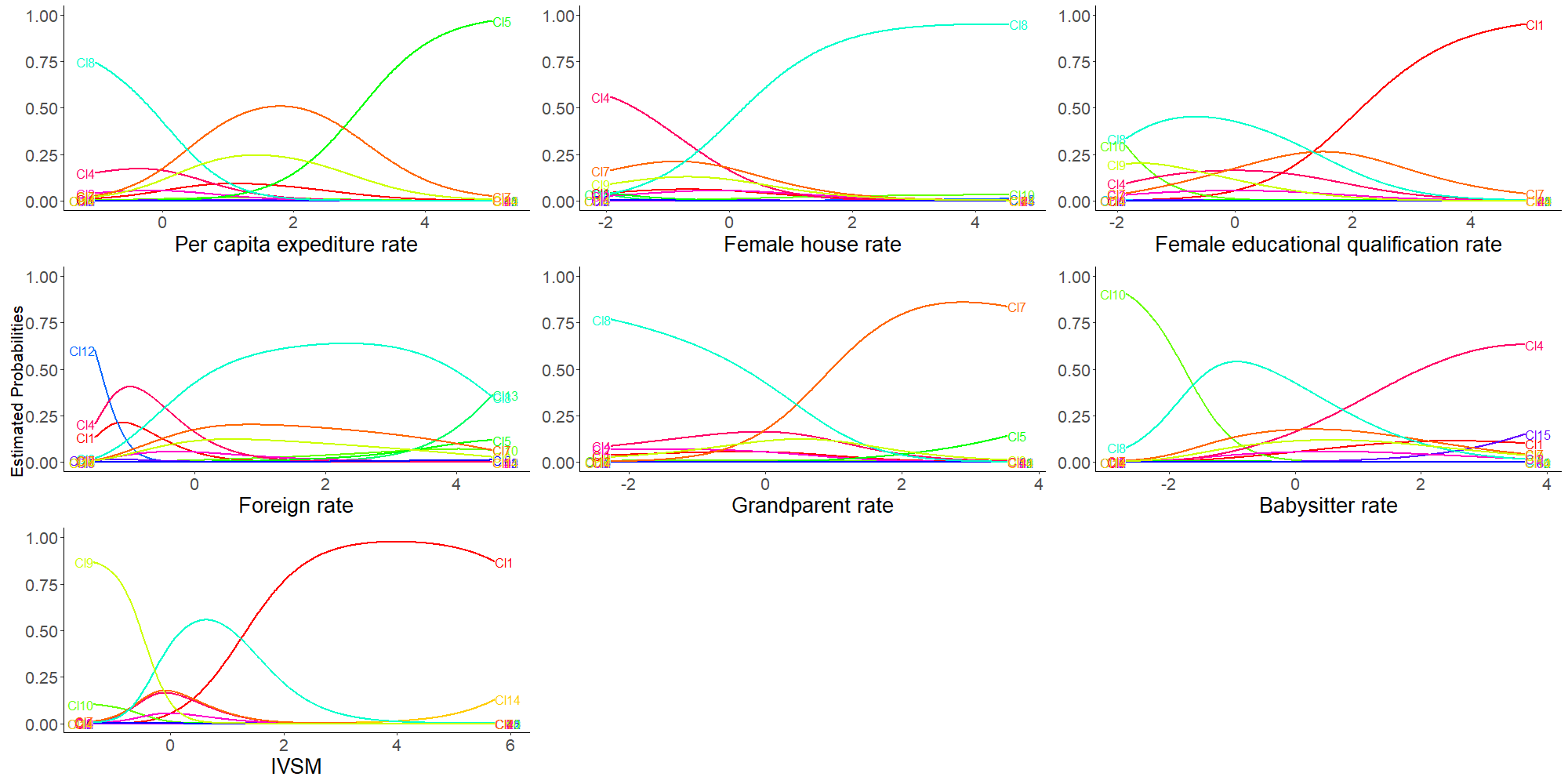}

   \caption{Predicted probabilities of being in clusters $j$, where $j = 1, \dots, 15$ across several values of 
 per capita expenditure rate (left top figure), female home rate (center top figure), female educational qualification rate (right top figure), foreign rate (left middle figure), grandparent rate (center middle figure),  babysitter rate (right middle figure), and IVSM (left bottom figure).}\label{fig:multinom2}
\end{figure}

Table \ref{tab:multinom_glmnet} shows the coefficients estimated from the penalized multinomial model discussed in Section \ref{results}, while Figure \ref{fig:multinom2} represents the predicted probabilities in the same way as Figure \ref{fig:multinom} for the remaining variables described in Table \ref{tab:var}.

\end{document}